%% file: main.tex
\documentclass[10pt, conference, letterpaper]{IEEEtran}

\ifCLASSINFOpdf
\else
\fi

\usepackage{fancyhdr}
\usepackage[pdftex]{graphicx}
\usepackage[hyphens]{url} 
\usepackage{xcolor}  
\let\labelindent\relax
\usepackage{enumitem}
\usepackage[breaklinks=true]{hyperref}         %
\hypersetup{                    %
	colorlinks,
	linkcolor={green!80!black},
	citecolor={red!70!black},
	urlcolor={blue!70!black}
}
\usepackage{breakurl}  
\usepackage{cleveref}
\usepackage{etoolbox}
\usepackage{multirow}
\usepackage{multicol}
\usepackage{float}
\usepackage{adjustbox}
\usepackage{tabularx}
\usepackage{longtable}
\usepackage[normalem]{ulem}
\usepackage{soul,xcolor}
\setstcolor{blue}

\input{macros}

\begin{document}

\iftoggle{revision}{
	\pagenumbering{gobble}
	\input{sections/revision-log}
}

\renewcommand{\paragraph}[1]{\textbf{#1.}}
\title{Bare Demo of IEEEtran.cls for Conferences}
\title{\Large \bf \sys: On-Demand and Scalable\\Cache Isolation for Security Architectures}

\iftoggle{camera}%
{\author{
\rm{Ghada Dessouky, Alexander Gruler, Pouya Mahmoody, Ahmad-Reza Sadeghi, Emmanuel Stapf}\\
Technische Universit\"at Darmstadt, Germany\\
\normalsize{\{ghada.dessouky, pouya.mahmoody, ahmad.sadeghi, emmanuel.stapf}\}\normalsize{@trust.tu-darmstadt.de}}
}
{\author{
}} %

\maketitle

\begin{abstract}\label{sec:abstract}
Shared cache resources in multi-core processors are vulnerable to cache side-channel attacks. %
Recently proposed defenses such as randomized mapping of addresses to cache lines or well-known cache partitioning have their own caveats: Randomization-based defenses have been shown vulnerable to newer attack algorithms besides relying on weak cryptographic primitives. They do not fundamentally address the root cause for cache side-channel attacks, namely, mutually distrusting codes sharing cache resources. %
Cache partitioning defenses provide the strict resource partitioning required to effectively block all side-channel threats. However, they usually rely on way-based partitioning which is not fine-grained and cannot scale to support a larger number of protection domains, e.g., in trusted execution environment (TEE) security architectures, besides degrading performance and often resulting in cache underutilization.

To overcome the shortcomings of both approaches, we present a novel and flexible set-associative cache partitioning design for TEE architectures, called \sys. The core idea of \sys is to enable an execution context to ``carve'' out an exclusive \emph{configurable chunk} of the cache if the execution requires side-channel resilience. If side-channel resilience is not required, mainstream cache resources can be freely utilized. Hence, our solution \sys addresses the security-performance trade-off practically by enabling efficient selective and on-demand utilization of side-channel-resilient caches, while providing well-grounded future-proof security guarantees. We show that \sys provides side-channel-resilient cache utilization for sensitive code execution, with \CHANGED{small} hardware overhead, while incurring no performance overhead on the OS. We also show that it outperforms conventional way-based cache partitioning by 43\%, while scaling significantly better to support a larger number of protection domains.

\end{abstract}

\input{introduction}

\input{preliminaries} %
\input{system-attacker} %
\input{design} %
\input{security} %
\input{eval} %
\input{related} %
\input{conclusion}

\appendix
\input{app-eval}

{\bibliographystyle{plain}
	{\footnotesize
		\bibliography{main_bib}}}

\end{document}

%% file: macros.tex
\newtoggle{draft}
\settoggle{draft}{false}

\newtoggle{revision}
\settoggle{revision}{false}

\newtoggle{confidential}
\settoggle{confidential}{false}

\newtoggle{camera}
\settoggle{camera}{true}

\newcounter{todos}

\newcommand{\CHANGED}[1]{\iftoggle{revision}{\textcolor{blue}{#1}}{#1}}

\usepackage{xspace}
\newcommand{\sys}{\textsc{Chunked-Cache}\xspace}
\newcommand{\sd}{I-Domain\xspace}
\newcommand{\sds}{I-Domains\xspace}
\newcommand{\nsd}{NI-Domain\xspace}

\newcommand{\idid}{\ensuremath{\mathtt{DID}}\xspace}
\newcommand{\shared}{\ensuremath{\mathtt{SHARED}}\xspace}

\newcommand{\etable}{\textsc{DCAT}\xspace}
\newcommand{\etablelong}{\textsc{Domain Cache Allocation Table}\xspace}
\newcommand{\cstlong}{\textsc{Cache Set Status Table}\xspace}
\newcommand{\cst}{\textsc{CST}\xspace}
\newcommand{\dcache}{\textit{cache chunk}\xspace}
\newcommand{\dcaches}{\textit{cache chunks}\xspace}
\newcommand{\mode}{\textsc{Isolation Mode}\xspace}
\newcommand{\modes}{\textsc{Isolation Modes}\xspace}

\newcommand{\nonisolated}{\textsc{Mainstream-Cache Mode}\xspace}
\newcommand{\isolated}{\textsc{Exclusive-Cache Mode}\xspace}

\newcommand{\mainstreammode}{\textsc{Mainstream-Cache Mode}\xspace}
\newcommand{\exclusivemode}{\textsc{Exclusive-Cache Mode}\xspace}
\newcommand{\sharedmem}{\textsc{Shared Memory}\xspace}

%% file: introduction.tex
\section{Introduction}
\label{sec:intro}
The outbreak of micro-architectural attacks has demonstrated the crucial implications of performance-boosting processor optimizations on the security of our computing platforms~\cite{Kocher18,google-zero, Lipp18, koruyeh2018spectre, Kiriansky18, maisuradze2018ret2spec, cachebleed2017yarom, gras2018translation, evtyushkin2018branchscope, evtyushkin2016jump,lee2017inferring,aciiccmez2007predicting, aciiccmez2007power, clkscrew2017tang, memjam2018moghimi, foreshadow, ridl, fallout, sgxpectre, zombieload, lvi}. %
One of the most popular features, and also the subject of many recent attacks, are shared resources such as caches. Caches provide orders-of-magnitude faster memory accesses and large last-level-caches (LLCs) are usually shared across multiple processor cores to maximize utilization. 

\paragraph{The Problem with Caches} 
When a sensitive (victim) and malicious (adversary) application run simultaneously on different cores and share the LLC, cache side channels can be exploited by the adversary to leak sensitive information, such as private keys. The timing difference between a cache hit and miss -- which is why caches are used in the first place -- is the most commonly exploited side channel to infer the memory access patterns of a victim application~\cite{Gullasch11, Yarom14, Gruss16, Irazoqui16, Lipp16, Irazoqui15, Kayaalp16, Liu15, Gras2017, Osvik2006, Gruss15, Guanciale16, Yan19, VanSchaik18}. In typical side-channel attacks~\cite{Osvik2006, Irazoqui15, Kayaalp16, Liu15, Gullasch11,Yarom14} the adversary deduces the victim's memory access patterns by exploiting that both the victim and adversary compete for shared set-associative cache resources, which are designed in such a way that a larger number of memory lines are mapped to a smaller number of cache ways/entries in each cache set.

Besides compromising cryptographic implementations~\cite{bernstein2005cache, Liu15, Osvik2006, Yarom14}, more recent attacks have had even stealthier impact such as bypassing address space layout randomization (ASLR) or leaking privacy-sensitive human genome indexing computation~\cite{Lipp16, Gras2017, Brasser17, Gruss16, Gruss15}, leaving millions of platforms vulnerable. Even trusted execution environment (TEE) security architectures which aim to protect sensitive services by compartmentalizing them in isolated execution contexts, called \textit{enclaves}, e.g., Intel SGX~\cite{intel-sgx1, intel-sgx2} or ARM TrustZone~\cite{trustzone}, have been shown vulnerable to these attacks, thereby undermining their acclaimed privacy and isolation guarantees~\cite{Brasser17,Schwarz17, Moghimi17, GES17, ARMageddon,Zhang16}.
This is alarming since TEE architectures are now widely deployed by major cloud providers, e.g., Microsoft Azure, Google Cloud, Alibaba Cloud and IBM Cloud, to offer \textit{confidential computing}, where sensitive workloads are protected in enclaves.

\paragraph{The Problem with Recent Cache Defenses}
To mitigate cache side-channel attacks, various approaches have been proposed over the years. %
These solutions range from time-constant 
cryptographic implementations~\cite{Doychev15,Doychev17,Kopf12} to software- and hardware-based approaches that modify the cache organization itself. 
The latter can be broadly classified into either cache partitioning~\cite{Godfrey03,Wang16,Kiriansky17,Liu16, hybcache,Kiriansky17} or randomization-based techniques~\cite{Newcache16, Trilla18, Qureshi18,ceaser-s, scattercache, phantomcache} that attempt to obfuscate the relationship between the memory address and the cache location \CHANGED{to which it is mapped.}

More recently, various schemes for a randomized memory-to-LLC mapping, such as CEASER, ScatterCache, and Phantom-Cache~\cite{Trilla18, Qureshi18,ceaser-s, scattercache, phantomcache} have been proposed to mitigate these attacks by
obfuscating the adversary's view of which cache lines actually get evicted. 
However, such defenses continue to evict cache lines from a small number of locations in a shared cache, thus cache set-based conflicts essentially still occur.
While these defenses were shown effective against the eviction set construction algorithms and techniques at the time, subsequent more efficient eviction set construction algorithms~\cite{ceaser-s} were able to undermine them. Consequently, enhancements to these defenses were proposed~\cite{ceaser-s}, only to be rendered ineffective again by yet another 
attack vector, e.g., weak low-latency cryptographic primitives~\cite{systematic-gruss, brutus}, or \CHANGED{alternative attack techniques that exploited design/implementation flaws in the proposed defenses~\cite{song-randomized}}. 

Caught in an arms race, randomization-based defenses remain as good as the best known attack technique at the time and are constructed to mitigate very specific side channels and attack strategies~\cite{casa-micro20}, with no future-proof and well-grounded security guarantees. They only make the attacks computationally more difficult, but do not address their fundamental root cause, i.e., sharing set-associative caches across mutually distrusting processes. %
These schemes also assume that all execution contexts require side-channel resilience without providing mechanisms for a selective configuration of side-channel-resilience, 
thus, taxing the entire system with the resulting performance impact. In practice, however, only a small portion of the workload is usually security-/privacy-sensitive and requires this sophisticated security guarantee.

On the other hand, strict partitioning approaches promise well-grounded security guarantees due to their cache isolation across different execution contexts. However, these approaches usually rely on conventional way-based partitioning~\cite{bahmani2020cure,keystone,Wang16,Kiriansky17,hybcache}, and thus, are not fine-grained, cannot scale with an increasing number of execution contexts and large LLCs, or do not provide support for shared memory. 

With these limitations in mind, we argue that a more future-proof and practical approach for side-channel resilient cache computing is to address the root cause of these attacks, namely, sharing set-associative cache structures across mutually distrusting execution contexts. Meanwhile, performance, usability, flexibility and scalability should still be preserved. We further observe that, in practice, \CHANGED{cache side-channel resilience is most prominently a concern in dedicated security architectures}, e.g., TEE security architectures. Thus, it is crucial to develop side-channel-resilient cache designs that cater for the security/functionality requirements of these architectures, e.g., with integrated support for enabling the side-channel resilience (and the performance cost) only for specific execution contexts that require it.

\paragraph{Our Goals}
In this work, we aim to selectively enforce clean partitioning of the cache resources across mutually distrusting execution contexts that require side-channel resilience, such that all side channels are blocked (including stealthy cache occupancy channels~\cite{WebFingerprinting} which are not mitigated by recent works~\cite{scattercache,hybcache}), while maintaining the desired performance requirements. 

To address this performance-security trade-off, we propose a new cache design for TEE security architectures, which we call \sys, that enables each execution context or domain to ``carve'' out its exclusive cache \emph{sets}, if desired. These sets essentially constitute an independent set-associative cache, which we call the domain's \dcache, that this domain can utilize exclusively but fully and efficiently, unlike in \CHANGED{cache partitioning, e.g., way-based partitioning}. A domain can flexibly request and configure 1.)~whether it requires side-channel-resilient cache utilization, 2.)~for which memory regions, and 3.)~the required capacity of this exclusive side-channel-resilient \dcache.
Memory accesses by a domain that requires side-channel-resilient cache utilization are mapped exclusively to its \dcache, while mainstream cache resources are freely and conventionally utilized whenever side-channel-resilience is not required. Enabling this on-demand flexibility per domain \emph{practically} requires addressing multiple key challenges. Firstly, efficient design mechanisms are required to configure the memory-to-set mapping at run time for each domain depending on its chunk capacity, while preserving conventional cache behavior for the rest of the execution. Secondly, it must be ensured that the operating system performance is not degraded as cache sets get allocated exclusively to domains. Finally, seamless support must be provided for shared memory between domains to meet the security and functionality requirements of different sensitive applications.

\paragraph{Our Contributions} Our main contributions are as follows:
\begin{itemize}[noitemsep,topsep=0pt]
	  \item We present \sys, a novel cache architecture for TEE security architectures, which enables a selective, flexible and scalable configuration of side-channel resilient caches for execution domains, without degrading the OS performance.
	  \item We address the performance-security trade-off by enforcing clean cache partitioning that blocks all cache side channels by allocating exclusive cache chunks for different domains. In doing so, future-proof and solid security assurances are guaranteed, while still preserving performance, functionality and compatibility requirements.
	  \item We extensively evaluate the cycle-accurate performance overhead of \sys for compute-intensive SPEC CPU2017 workloads and I/O-intensive real-world applications. We show that it outperforms shared cache utilization in some cases, that the OS performance even improves owing to \sys's flexible cache utilization, and that \sys outperforms \CHANGED{partitioning (way-based)} by 43\% while also scaling better to support a larger number of protection domains.
		\item We implement and evaluate a hardware prototype of \sys. We show that it incurs a minimal 2.3\% memory overhead relative to a 16 MB LLC, 1.6\% logic overhead relative to a single-core RISC-V processor, and 12.3\% LLC power consumption overhead.
\end{itemize}

%% file: preliminaries.tex
\section{Cache Attacks \& Defenses}
\label{sec:prelim}
Next, we briefly introduce recent cache side-channel attacks that are relevant for our work and a summary of the shortcomings of recent defenses that our work overcomes. %

\subsection{Cache Side-Channel Attacks} 
\label{sec:attacks-prelim}
Cache side-channel attacks have been shown to constitute a profound threat that underlies popular attacks such as Spectre~\cite{Kocher18} and Meltdown~\cite{Lipp18}, besides threatening a wide spectrum of platforms and architectures~\cite{ARMageddon,Liu15,Irazoqui15,Zhang12}, and even TEE architectures~\cite{Brasser17,Schwarz17, Moghimi17, GES17,ARMageddon,Zhang16}.
The attacks usually work by provoking controlled evictions of the victim's cache line, such that the inherent information leakage from the access-timing difference between cache hits and misses can be exploited by the adversary. This can be achieved using three main approaches:
\begin{itemize}[noitemsep,topsep=0pt]
	  \item Access-based approaches where the target address is explicitly accessed and flushed\CHANGED{~\cite{Gullasch11,Yarom14,Gruss16,Irazoqui16,Lipp16}}.
		\item Conflict-based approaches where the adversary triggers a controlled cache contention in the same cache set of the target address to evict the corresponding victim cache lines\CHANGED{~\cite{Osvik2006,Irazoqui15, Kayaalp16, Yarom14, Liu15, Yan19, Disselkoen17, Gras2017, Osvik2006, Guanciale16,VanSchaik18,bernstein2005cache, Bonneau06}}.
		\item Occupancy-based approaches~\cite{WebFingerprinting} where the adversary observes an eviction of its own cache lines and uses this information to infer the size of the victim's working set.
\end{itemize}

\subsection{Recent Defenses and their Shortcomings} 
\label{sec:defenses}
Various defenses against side-channel attacks have been proposed, focusing on access-based and conflict-based attacks.

\paragraph{Side-channel Resilient Implementation} This aims at implementing algorithms, e.g. cryptographic algorithms, in a time-constant (thus side-channel-resilient) fashion~\cite{into_crypto_alog,bernstein2012security}. Time-constant algorithms vary between hardware platforms~\cite{cock2014last} and require considerable effort that is not generalizable and scalable for all software.

\paragraph{Attack Detection}
Other approaches aim to detect attacks in progress by observing hardware performance counters (e.g., on cache miss rates)~\cite{chiappetta2016real,payer2016hexpads} and killing the suspicious process. However, being based on heuristics, attacks can only be discovered with a certain probability and no guaranteed protection is provided. Moreover, some attacks have been shown to not cause an abnormal cache behavior~\cite{Gruss16}.

\paragraph{Noisy Measurements} Another group of defenses aims to impede a successful attack by preventing the adversary from performing precise time measurements, e.g., by restricting the access to timers~\cite{osvik2006cache,percival2005cache,martin2012timewarp}, by injecting noise into the system~\cite{vattikonda2011eliminating,hu1992reducing} or deliberately slowing down the system clock~\cite{Hu91,Martin12}. However, workarounds have been found to create timers~\cite{schwarz2017fantastic} or to perform attacks without relying on timers~\cite{disselkoen2017prime}. Moreover, such defenses cannot protect TEE architectures since they assume a strong adversary that can compromise the OS kernel and circumvent such restrictions.

\paragraph{Cache-level Defenses} 
Other approaches tackle the side-channel problem directly where it originates, i.e., at the cache level.
These defenses fall under one of two paradigms: 1.)~randomized cache line mapping
to make the attacks computationally impractical~\cite{Trilla18,Qureshi18,ceaser-s,scattercache,phantomcache,Wang07,Newcache16,Liu14} or 2.)~cache partitioning to provide strict isolation~\cite{Godfrey03,Kim12,Zhang16,Liu16,costan2016sanctum,Gruss17,zhao2019sectee,Kayaalp17,Yan17,keystone,bahmani2020cure,Wang16,Kiriansky17,Wang07,hybcache}.
We discuss the works most related to \sys in more detail in~\Cref{sec:related}.

Randomization-based defenses cannot provide comprehensive future-proof security guarantees, e.g., advances in attack strategies and minimal eviction set construction techniques, \CHANGED{besides alternative attack techniques have been shown to undermine such defenses~\cite{ceaser-s,casa-micro20,purnal2019advanced,systematic-gruss,song-randomized}.}
Moreover, many rely on cryptographic primitives which have been shown vulnerable to cryptoanalysis, while deploying more secure primitives would further degrade performance~\cite{brutus,systematic-gruss}.

Cache partitioning defenses provide strict resource isolation which allows to give solid security guarantees on side-channel protection. However, existing partitioning defenses suffer from high performance penalties, restrictive and inflexible cache utilization~\cite{Wang07} and their inability to scale with a larger number of protection domains~\cite{Wang16,Kiriansky17,Gruss17}. Several approaches do not directly cater for the use of shared libraries~\cite{Godfrey03,Wang16}, are architecture-specific~\cite{Kayaalp17, Yan17} or do not defend against occupancy-based attacks. Memory page coloring approaches~\cite{costan2016sanctum, Kim12,Godfrey03} are impractical since they require invasive modifications of the memory management of commodity software and cannot sufficiently support Direct Memory Access (DMA). Most importantly, existing partitioning defenses to date apply their side-channel cache protection for the entire execution workload, impacting overall system performance, which is not even required in most scenarios.

To fundamentally address all these shortcomings, we propose a modified cache microarchitecture, which we call \sys, that provides strict, yet configurable partitioning across the mutually distrusting execution domains. For each domain, \sys carves out and isolates an exclusive cache share only as the domain requires. %
This effectively mitigates all interference across domains, thus, defending against even stealthy cache occupancy attacks unlike recent cache defenses, while activating side-channel resilience only for sensitive execution domains that require it. All other execution domains can freely utilize mainstream cache resources at the same performance or even improved performance than conventional non-secure cache sharing.

%% file: system-attacker.tex
\section{System \& Adversary Model}

In the following section, we describe our assumptions regarding the system and adversary model. 

\subsection{System Model}
\label{subsec:system_model}

\sys targets computing systems which implement a \CHANGED{TEE security architecture} and contain a set-associative cache architecture. In the following, we first present our standard assumptions regarding the cache architecture, followed by our assumptions on the TEE security architecture which are aligned with existing academic~\cite{costan2016sanctum,keystone,sanctuary,bahmani2020cure} and industry solutions~\cite{intel-sgx1,amd_sev,trustzone}.

\paragraph{Cache Architecture} 
In \sys, we assume a typical modern set-associative cache architecture with multiple cache levels, where some cache levels are core exclusive (typically L1 and L2) and others shared between multiple cores (L3), whereby the L3 can be a sliced cache, e.g., sliced Intel LLCs.
While \sys can be deployed to provide partitioning for smaller L1 and L2 caches in principle, we assume, however, that core-exclusive caches are flushed at context switching (similar to most recent TEE architectures~\cite{bahmani2020cure, costan2016sanctum, keystone}), and thus, that \sys is deployed for the last-level L3 cache. Moreover, we assume that the cache controller can be configured via dedicated configuration registers, in line with typical platforms. 

\paragraph{TEE Architecture}  
We assume that the computing systems which deploy \sys implement a TEE architecture. TEE architectures already have established mechanisms for protecting sensitive code in compartmentalized execution contexts called \textit{enclaves} or \textbf{I}solated \textbf{D}omains (\sd{}), as we refer to them in this work. All non-sensitive code which does not require enhanced protection is consolidated in a \textbf{N}on-\textbf{I}solated \textbf{D}omain (\nsd{}). The domains are also each assigned a unique identifier (domain ID). The separation between the \sds and the \nsd is enforced by access control mechanisms already implemented in the TEE architectures, e.g., at the MMU in Intel SGX~\cite{intel-sgx1} or Sanctum~\cite{intel-sgx1}, at the system bus in CURE~\cite{bahmani2020cure} or by the Physical Memory Protection (PMP) unit in Keystone~\cite{keystone}. The access control mechanisms are either configured by microcode~\cite{intel-sgx1,amd_sev} or by a small software component which consists only of a few thousand lines of code (to be formally verifiable) and which runs in the highest software privilege level of the system~\cite{costan2016sanctum,keystone,sanctuary,bahmani2020cure,trustzone}. We refer to this component as a \textit{trusted software component}. The trusted software component is also responsible for all other security-sensitive operations, e.g., assigning the domain IDs, and, in the case of \sys, configuring our novel protection mechanisms in the cache controller which we describe in detail in~\Cref{sec:design}.

Although \sd{}s are security-sensitive, they might still require to share data with another domain, e.g., to enable communication with the operating system. Thus, TEE architectures typically provide the possibility to mark parts of an \sd{}'s memory as \textit{shared}, whereby this information is again managed by the trusted software component. In many TEE architectures, e.g., TrustZone~\cite{trustzone}, CURE~\cite{bahmani2020cure} or AMD SEV~\cite{amd_sev}, security-relevant metadata, which is required to perform access control, is sent as part of every memory request. For \sys we assume the same, namely, that the domain ID of the domain issuing a memory access request and the information whether the requested memory address is \textit{shared} or \textit{non-shared}, are sent within the memory request.

\subsection{Adversary Model}
\label{subsec:attacker_model}
Since we focus on the deployment of \sys on systems with TEE architectures, we assume the same strong adversary model where the operating system kernel and hypervisor are untrusted~\cite{costan2016sanctum,keystone,sanctuary,bahmani2020cure,intel-sgx1,amd_sev,trustzone}.

With regard to cache side-channel attacks, we assume the adversary \CHANGED{has access to the \sys specification and} is able to mount access-based and conflict-based side-channel attacks, which are the most sophisticated and applicable cache attacks (cf.~\Cref{sec:attacks-prelim}), to leak information about a sensitive execution domain (\sd). Since the adversary is also able to control the OS kernel, we assume a worst-case scenario where an adversary can easily mount the described attacks, i.e., has knowledge about the \sys design and specs, and knows the virtual to physical address mapping of the victim domain. Moreover, the adversary can mount attacks from all privilege levels (except the highest privilege level that contains the trusted software component), has access to precise timing measurements and eviction instructions (e.g., \texttt{clflush}), can attack from the same CPU core executing the victim domain or a different core (cross-core), freely interrupt the victim domain and even keep the system noise to a minimum. In contrast to related work~\cite{Trilla18, Qureshi18,ceaser-s, scattercache, phantomcache}, we also consider the stealthier cache occupancy-based attacks (cf.~\Cref{sec:attacks-prelim}). Collision-based attacks~\cite{Bonneau06}, which exploit cache collisions at the victim caused by the victim's own cache utilization, are\CHANGED{, aligned with related work, kept out of scope. Collision-based attacks have not been widely shown and are very specific to particular software implementations (e.g., table-based).}

Apart from cache side-channel attacks, an adversary who compromises the OS kernel has full control over the memory management and thus, can easily map physical memory pages of a victim domain into its own memory. This allows an adversary to perform rogue cache accesses to sensitive data \textit{directly} without the need of a cache side channel.

In line with related work~\cite{Godfrey03,Wang16,Kiriansky17,Liu16, hybcache,Wang07,Newcache16, Trilla18, Qureshi18,ceaser-s, scattercache, phantomcache}, we do not consider physical attacks on caches, e.g., physical side-channel attacks~\cite{hw_side_channel}, fault injection attacks~\cite{fault_inject}, and attacks that exploit hardware flaws~\cite{clkscrew2017tang,kenjar2020v0ltpwn,qiu2019voltjockey}.
We do not consider denial-of-service attacks from a security point of view. However, to avoid the performance impact on the OS, \sys ensures that a certain amount of cache resources are always available to the OS (described in~\Cref{sec:design}).
Based on our system model (\Cref{subsec:system_model})
, we assume that the adversary cannot compromise the trusted software component.

%% file: design.tex
\section{Chunked-Cache Design} \label{sec:design}
We first describe the high-level idea of \sys, a novel cache microarchitecture that provides flexible and on-demand assignment of cache resources to execution domains (\Cref{subsec:high-level}). We follow with a detailed explanation of our design (\Cref{subsec:details}) and the required cache tag store and cache controller modifications (\Cref{subsec:controller}). 

\begin{figure}[ht]
	\centering
	\includegraphics[width=1\columnwidth]{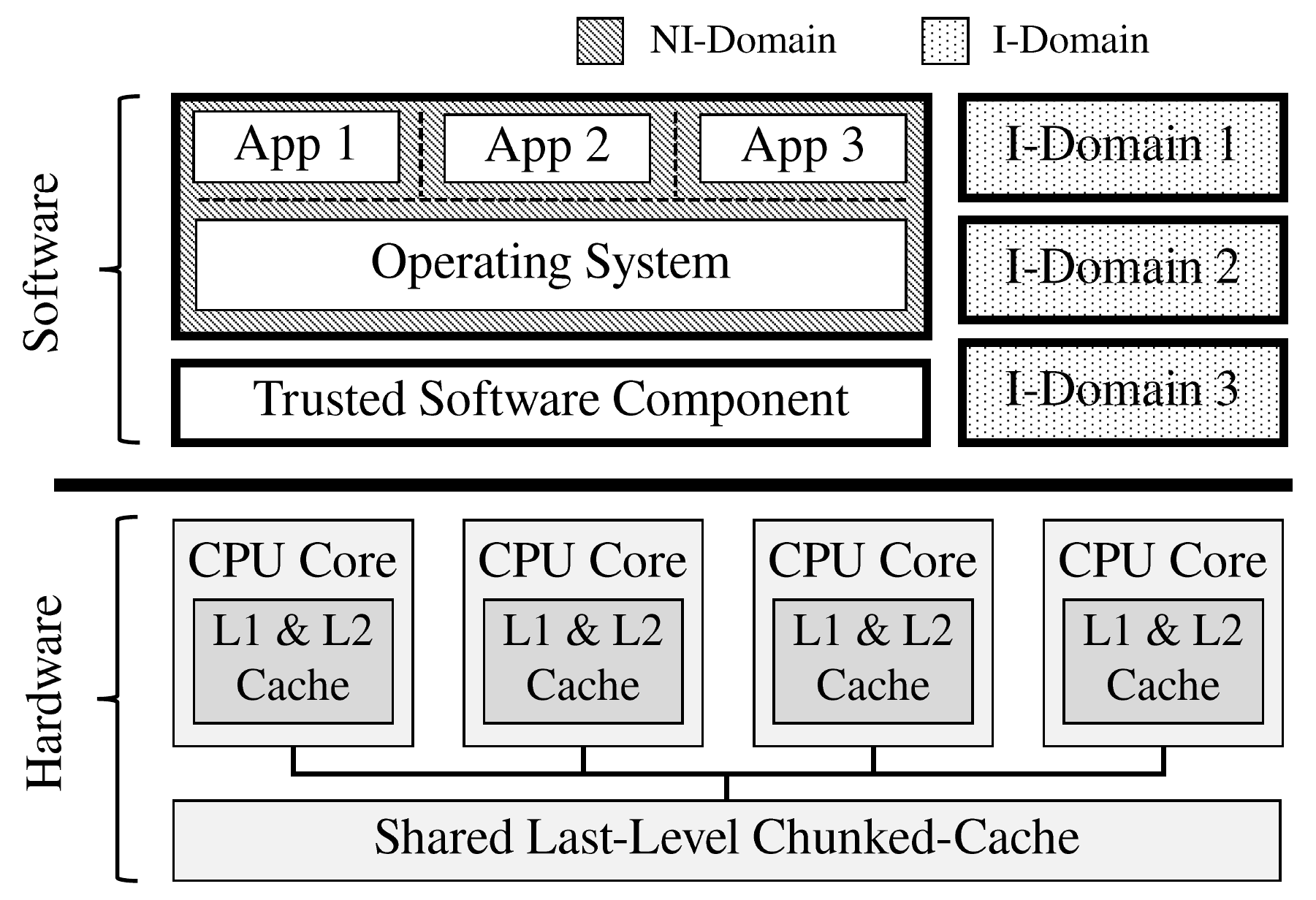}
	\caption{\small Computing system with TEE architecture and \sys as the shared last-level cache.}
	\label{fig:system-high-level}
\end{figure}

\subsection{High-Level Design} \label{subsec:high-level}

\begin{figure}[ht]
    \centering
    \includegraphics[width=1\columnwidth]{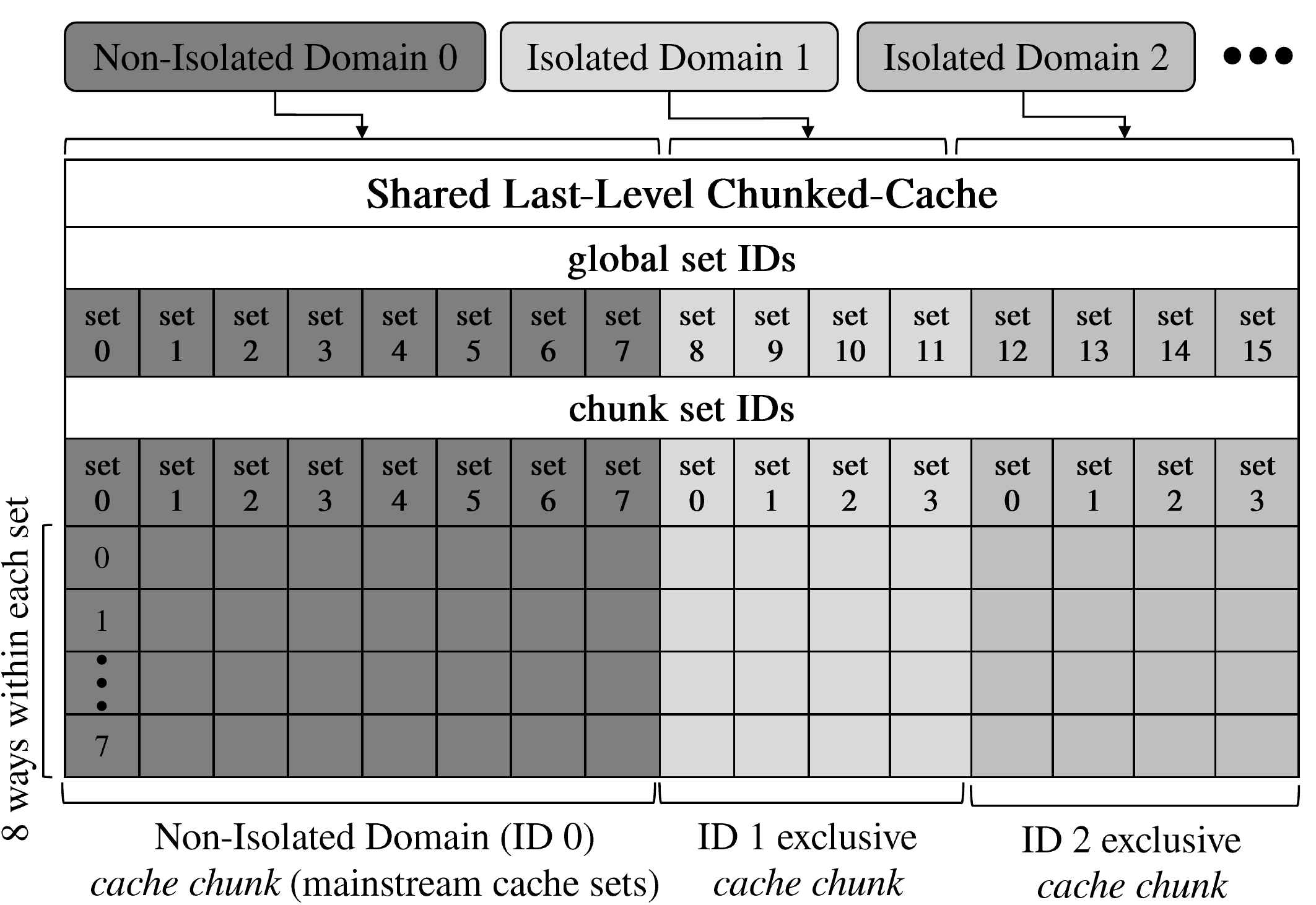}
    \caption{\small \sys high-level design: each domain gets an exclusive \dcache allocated on-demand.}
    \label{fig:high-level}
\end{figure}

In~\Cref{fig:system-high-level}, we show how \sys is integrated as the last-level cache in a computing system which implements a TEE architecture, aligned with our system model detailed in~\Cref{subsec:system_model}. \Cref{fig:high-level} steers the focus to the design of \sys itself and illustrates its architecture abstractly.
As described in~\Cref{subsec:system_model}, all TEE architectures provide built-in mechanisms to protect sensitive code in \textbf{I}solated \textbf{D}omains (\sd{}s), whereas non-sensitive code is running in a \textbf{N}on-\textbf{I}solated \textbf{D}omain (\nsd{}).

Each active domain (\nsd and \sd{}s) is uniquely identified by an ID: \idid. The operating system (OS) and all workloads which do not require protection (and are combined in the \nsd) are assigned the \idid 0 
by default. 
Every \sd{} can request exclusive cache resources of desirable capacity, forming the domain's exclusive \dcache, that is only utilized by the owner domain. %
The \nsd utilizes the cache sets which are not exclusively allocated to \sds, which we call \emph{mainstream} cache sets.

Each \sd requests its dedicated \dcache consisting of the required number of cache sets, e.g., \sd 1 in~\Cref{fig:high-level} requested 4 sets. 
Thus, at \sd 1 setup, 4 available (unallocated) sets are located in the cache (sets with global IDs 8-11 here) and allocated to \sd 1 such that they form its \dcache. The allocated sets are mapped to \sd 1's chunk set IDs 0-3, and they are used to exclusively cache all and only memory accesses issued by \sd 1.
Enabling each \sd to request its desired \dcache capacity exclusively provides strict partitioning and completely isolates its cache utilization on-demand. Besides enabling selective cache-based side-channel resilience, this also allows that each \sd acquires the performance that corresponds to the cache capacity it has requested, without any competition from other workload.
In contrast to \CHANGED{partitioning schemes~\cite{Liu16,keystone,bahmani2020cure,Wang16,Kiriansky17}} that provide each domain with only 1 or 2 ways within each set of the full cache structure, \sys also partitions the cache but more efficiently. \sys carves out a full \dcache (with all its ways per set) of configurable capacity for the \sd and configures all its memory accesses to be mapped to the \dcache, thus promising \CHANGED{maximum and unshared utilization of the allocated \dcache}. \CHANGED{We show in~\Cref{sec:eval} that \sys provides better performance and enhanced scalability than partitioning schemes.}

\CHANGED{By allowing each \sd a custom and configurable \dcache capacity on-demand, in contrast to fixed allocation, \sys enables an adaptive security-performance trade-off in the cache microarchitecture. On one hand, non-sensitive workload can be allowed to freely utilize the shared mainstream cache resources. On the other hand, if side-channel resilience is a concern, a \dcache with default capacity can be allocated to each \sd without any further intervention from the developer. Only if the developer requires to further optimize the performance of the workload in a particular \sd, then the \dcache capacity (its number of sets) can be accordingly calibrated, i.e., assigning an \sd more cache resources if affordable/available.}

\subsection{Design Details of \sys} \label{subsec:details}
In the following, we discuss the key design goals and challenges of \sys, and the mechanisms we propose to achieve them.

\paragraph{Configurable Per-Domain Isolation Modes}
One of our key design goals for \sys is to support configurable cache isolation modes that provide different security guarantees, thus catering for different use cases and their requirements. In line with the design paradigm of TEEs, it is not reasonable to assume that all workloads require cache isolation and side-channel resilience.
Thus, in \sys, we provide 2 different \modes that each \sd can selectively configure for the workload it protects: 1.)~\mainstreammode: where cache isolation and side-channel resilience is not a security requirement, and thus, the \sd can utilize the mainstream cache. However, the cached \sd data must still be protected from malicious OS accesses. 2.)~\exclusivemode: where cache isolation is required since side-channel resilience is a security requirement and thus, an exclusive \dcache is required by this \sd. The latter mode is configured for \sd 1 and \sd 2 shown in~\Cref{fig:high-level}. In addition to the \mode, the \sd can also configure its \sharedmem settings, i.e., if it requires to share memory regions (and thus cache lines) with the OS, e.g., when using OS services. To cache shared memory, the mainstream cache that the OS uses is utilized. Typically, the developer of the workload decides which \mode an \sd uses and identifies which memory regions need to be shared, which is on par with the requirement in TEE architectures where the developer must identify the security-sensitive parts of the overall workload~\cite{intel-sgx1,trustzone,costan2016sanctum}. If a developer is not sure whether cache side-channel attacks are a threat, the ~\isolated should be selected out of caution. At setup, an \sd configures: 1.) the desired \mode for its cache utilization and 2.) its \sharedmem regions if required. 
This metadata is securely configured by the trusted component (as shown in~\Cref{fig:system-high-level}). The \mode is communicated to the cache controller at domain setup, whereas the \sharedmem information is transmitted at every memory request, aligned with our assumed system model (\Cref{subsec:system_model}). 		

\paragraph{Mainstream Cache vs. Shared Memory Support}
When an \sd is in \mainstreammode, it uses the mainstream cache sets also used by the OS (\idid 0). 
To prevent a malicious OS from mapping the memory of an \sd in its own memory space and accessing it directly in the cache, \sys requires that cache lines are tagged with the domain ID \idid. The hardware mechanisms integrated into the \sys controller enforce this tagging when caching the data, and that only the owner domain which cached the data can access it. Being hardware managed, the OS has no means to modify the \idid stored in the cache lines.

When an \sd is also sharing memory with the OS, the corresponding cache lines for the defined \sharedmem regions are cached in the mainstream cache sets, and are to be accessed by both the owner domain and the OS. To support that, cache lines need to be tagged with an additional \shared flag that indicates whether the cache line is shared with the OS. For typical TEE architectures, the developer of the workload protected in the \sd configures which of its memory regions are to be shared.

\paragraph{\exclusivemode Chunk Set Indexing}
The \emph{index} bits of a memory address are used to locate the cache set \CHANGED{to which it is mapped.} In a conventional cache, the number of index bits is fixed and depends on the number of sets the cache supports. However, for \sys to support \dcaches of different sizes for different domains, \emph{configurable set indexing} is required.

When an \sd is in \exclusivemode and requests a number of cache sets for its \dcache, the number of set index bits that will be used to map its memory lines has to be computed individually for this domain. Therefore, the cache controller keeps track of the global IDs of sets which constitute the \dcache (\Cref{fig:high-level}), %
and the index bits for each domain. When a memory access is issued by a domain, this metadata is looked up, and the pertinent \dcache sets correctly indexed. 
Moreover, when an \sd is torn down and its sets are de-allocated, the relevant metadata needs to be updated accordingly, besides flushing and invalidating the cache lines. \sys also enables support for dynamic cache allocation, i.e., allocating additional cache sets to an \sd's \dcache at runtime and reconfiguring the index bits accordingly. 
In~\Cref{subsec:controller}, we describe how the cache microarchitecture and controller are modified to enable this configurability efficiently.

\paragraph{\nsd Chunk Set Indexing}
Another design challenge in \sys is managing the sets allocated to the OS, which represents the \nsd with \idid 0, such that both flexibility as well as maximum utilization \CHANGED{(as in an unmodified insecure cache architecture)} are preserved. 
At bootup, when no domains are set up yet besides the OS, the OS should ideally be able to utilize all the available cache capacity, i.e., all cache sets are allocated to the OS by default. We refer to these as the \emph{mainstream} cache sets. Then, once domains are set up and request exclusive cache sets, these get ``torn away'' from the OS's cache and are allocated to the domains. 
This would, however, incur an impractical performance degradation for the OS since every time some of the OS's cache resources are allocated to another domain, its own capacity is changed, and so would its set indexing. This renders all memory lines already cached by the OS inaccessible unless complicated remapping is performed. 
Essentially, the OS would need to cache these memory addresses once again, thus suffering a high number of cold misses every time a new domain is set up and subjecting the OS to an unreasonably high performance overhead.

\begin{figure}[ht]
    \centering
    \includegraphics[width=1.0\columnwidth]{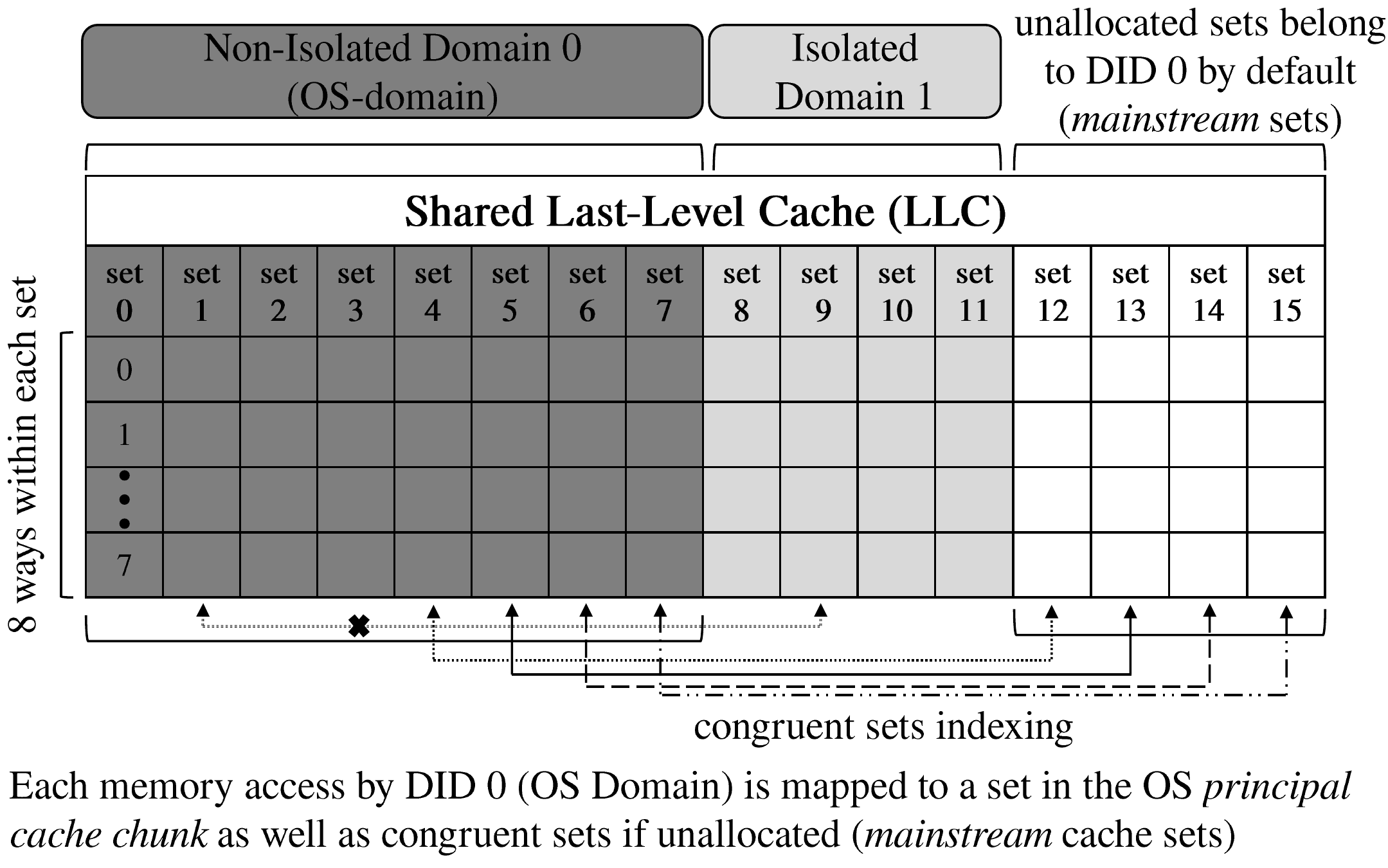}
    \caption{\small \sys OS-specific chunk set indexing.}
    \label{fig:os-chunks}
\end{figure}

To avoid this performance penalty on the OS, the OS is allocated a fixed (sufficiently large) number of the cache sets in \sys which remain always dedicated to the OS, while still allowing it to utilize the other cache sets so long as they remain unallocated. We demonstrate this in~\Cref{fig:os-chunks} where the OS is always allocated a fixed number of 8 sets (0-7) which form its principal \dcache.
Since the 8 sets are always available for the OS, the memory address indexing and the number of index bits do not change at runtime. In other words, no OS memory lines cached in this principal \dcache must ever be flushed out when any other domain requests to allocate additional cache sets, since the OS \dcache sets are never torn away from the OS. However, the OS can still utilize unallocated sets (sets 12-15) in parallel until they get allocated to another domain, thus also guaranteeing maximum utilization of the available cache resources. This works by indexing cache sets in parallel which are congruent to the set \CHANGED{to which a memory address is mapped.}
In~\Cref{fig:os-chunks}, 3 index bits are required to map a memory address to the correct set for a \dcache of size 8 sets. Thus, if the index bits, e.g., map to set 4, then set 12 can also be utilized by the OS (set ID + OS \dcache size) to cache that memory line. The same applies for memory lines that are mapped to sets 5, 6 and 7; they also map to sets 13, 14 and 15, respectively. However, memory lines mapped to sets 0-3 cannot utilize the congruent sets 8-11 because these are already allocated to \sd 1.

\begin{figure*}[ht]
    \centering
    \includegraphics[width=0.7\textwidth]{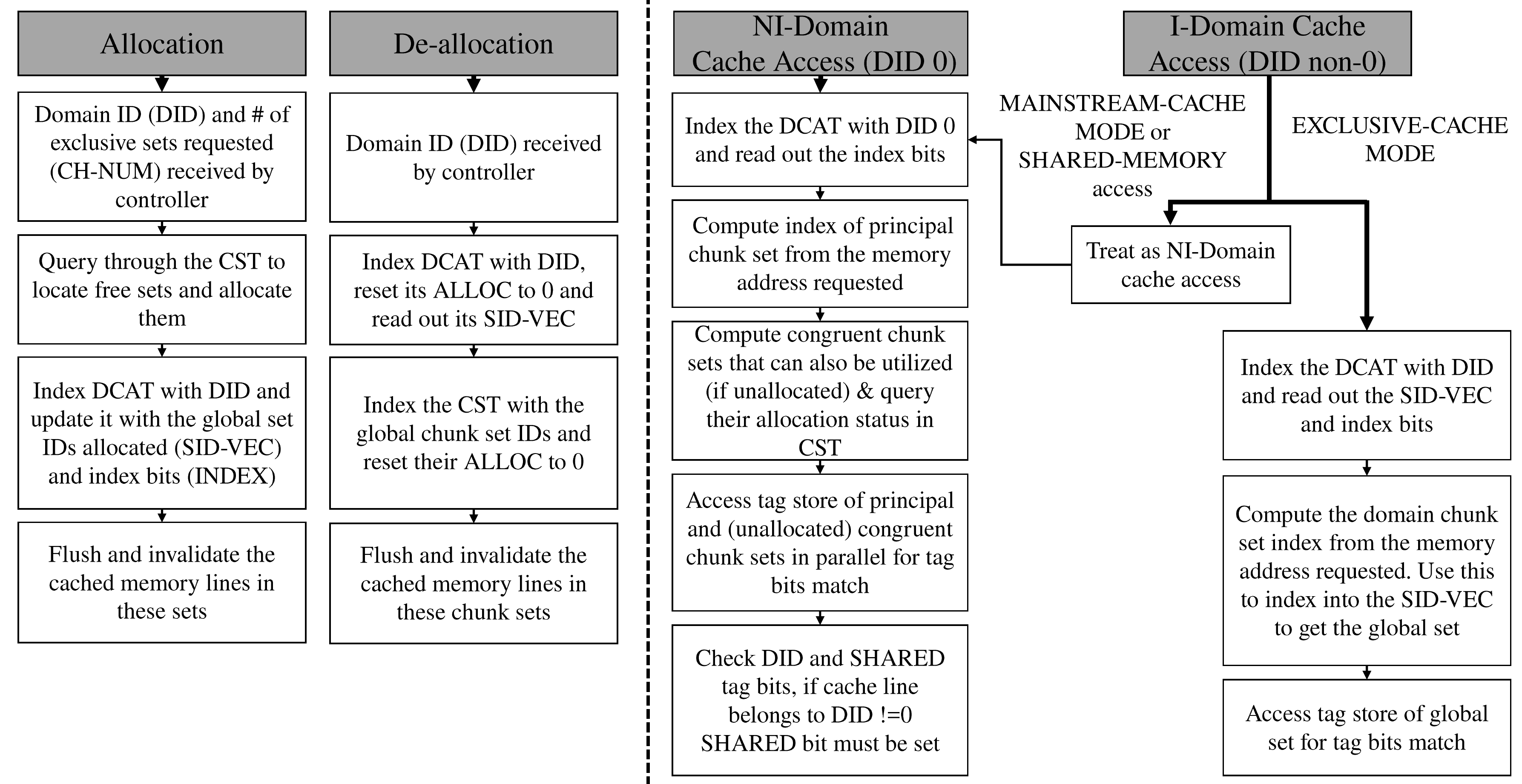}
    \caption{\small \sys controller operations for cache chunk allocation, de-allocation and access control.}
    \label{fig:control-flow}
\end{figure*}

\subsection{Cache Tag Store \& Cache Controller} \label{subsec:controller}
Cache lines need to be additionally tagged with the domain ID (\idid) bits as well as a 1-bit \shared flag bit to enforce access control and moderate sharing with the \nsd. For instance, to support 16 parallel active domains, we require to extend the cache tag store with 4 bits to represent the \idid.
We emphasize that the \sys design does not limit the number of parallel domains to 16; a larger number is possible but increases the hardware overhead of \sys (but only linearly). Moreover, the number of domains only limits how many domains can be simultaneously active on the system. It does not limit how many applications can be protected in \sds on the system in general. 

\begin{figure}[ht]
    \centering
    \includegraphics[width=1\columnwidth]{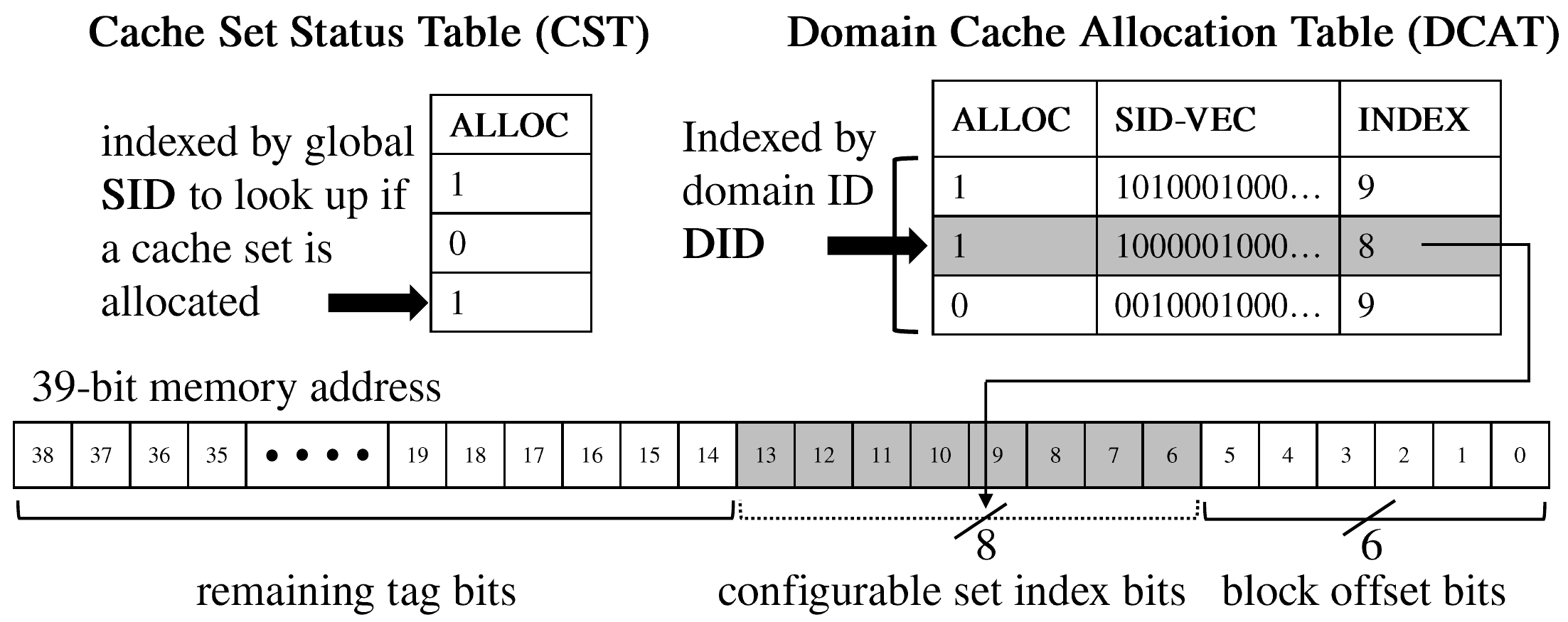}
    \caption{\small \sys table structures.}
    \label{fig:tables}
\end{figure}

To support the configurable set indexing, the allocation/de-allocation of cache sets to different \sds
and to differentiate between OS (\nsd) cache accesses vs. \sd accesses, 2 table structures are required by the \sys controller which are shown in~\Cref{fig:tables}. The \cstlong (\cst) is a 1-bit vector that is indexed by the global set ID (SID) and that stores the status of each set, i.e., whether it is allocated to a domain. The CST is used to query the status of a set %
when searching for free cache sets to allocate to an \sd.

The \etablelong (\etable) is indexed by the domain ID \idid. It maintains whether this domain is configured by the cache controller (\textsc{ALLOC}), a vector of the global set IDs that form its \dcache (\textsc{SID-VEC}), and the corresponding number of index bits (\textsc{INDEX}) required to map a memory line to the correct set ($log_2$(number of sets in the \dcache)), as shown in~\Cref{fig:tables}.

We describe next how the \sys controller performs these cache management operations, i.e., allocation, de-allocation and access control and represent this in~\Cref{fig:control-flow}. The description in~\Cref{fig:control-flow} only represents the sequence of operations for understanding, but does not reflect the temporal nature of the operations, i.e., whether they occur sequentially or in parallel.

\paragraph{Cache Allocation \& De-allocation} When an \sd requests to allocate exclusive cache sets, this request (\idid, the number of sets (\textsc{CH-NUM}) requested, and the corresponding number of \textsc{INDEX} bits ($\log _{2}$ \textsc{CH-NUM}) is securely communicated from the trusted component to the cache controller via configuration registers of the cache controller (\Cref{subsec:system_model}). The \idid is looked up in the \etable to check if it is already allocated and \CHANGED{that the maximum sets number allowed per \sd is not exceeded. The maximum/minimum limits for \sds are configured by the trusted software component, while ensuring that each \sd is always assigned at least a minimum \dcache size.}

The \cst is queried to locate free sets and to allocate them to the \sd by flipping the \textsc{ALLOC} bit, until \textsc{CH-NUM} sets are allocated. If \cst runs out of free sets, this is communicated back to the trusted component in order to modify the cache request. 
Next, the \etable is indexed with the \idid and its metadata updated by updating the \textsc{INDEX} bits and the \textsc{SID-VEC} with the global IDs of the allocated sets.

If a domain requests to de-allocate its cache sets, \etable is indexed with \idid, \textsc{ALLOC} reset and the \textsc{SID-VEC} read out. Next, the \cst is indexed with each set ID in \textsc{SID-VEC} and de-allocated. For both allocation and de-allocation, the cached memory lines in the relevant sets are invalidated and flushed (if dirty) to remove potentially malicious data in the allocation case and prevent information leakage in the de-allocation case.

\CHANGED{The number of cache sets which are always assigned to the \nsd are hardwired, since the circuitry for the parallel tag lookup (described below) must be hardwired and cannot be configurably extended.}

\paragraph{Cache Access Management} The \idid of an incoming cache access request indicates whether it is an access by the \nsd (OS domain with \idid 0) or an \sd. \CHANGED{If it is an OS access, then the index bits are fixed, since its number of cache sets are hardwired (no need to look its \textsc{INDEX} bits up in the \etable). The OS domain is assigned the least significant cache sets by default, thus the \textsc{SID-VEC} is also not needed.} The correct set index in the principal chunk is computed from the memory address in the request. Because it is an OS access, congruent cache sets that are not allocated can also be utilized (see~\Cref{subsec:high-level}). Thus, they are also computed and their \textsc{ALLOC} status queried in the \cst to locate the unallocated sets. \CHANGED{The tag store of the ways in the principal as well as the congruent sets are looked up in parallel to locate a tag bit match (cache hit), thus, neither impacting performance nor routing delay especially since a large number of principal sets are usually allocated to the \nsd which minimizes the number of congruent sets that are looked up in parallel (1 or 2 more sets).} The \idid and \shared tag bits are also checked in parallel. If the cache line belongs to a non-zero \idid (\sd), the \shared tag bit should be 1 to allow the OS to access it.

For an \sd (non-zero \idid), if access is requested to a \sharedmem region or if the \sd is in \nonisolated, then the access is treated by the controller as a \nsd access where the mainstream and congruent cache sets are accessed. However, at the tag comparison, the issuing \idid is checked against the cache line \idid to verify that only the owner domain accesses it. If the access is performed in \isolated, the exclusive \dcache of the domain is accessed. The \etable is indexed with the \idid and the %
\textsc{SID-VEC} and \textsc{INDEX} bits are read out. The chunk set index is computed and used to index into the \textsc{SID-VEC} to map to the correct global set ID. Then, the tag store is accessed for a tag bits comparison.

\sys's design is independent from the implemented cache replacement policy and thus, does not require additional modifications to it.
On every cache miss experienced by an \sd in \isolated, a cache line in the corresponding set in the domain's exclusive \dcache is selected for eviction. On cache misses by an \sd in \nonisolated or when accessing \sharedmem, and for all misses by the \nsd, a cache line in the corresponding set from the mainstream cache is selected.

%% file: security.tex
\section{Security Considerations}
\label{sec:security}
In this section, we discuss how \sys protects from the adversary described in~\Cref{subsec:attacker_model}. One key aspect of \sys is that its protection capabilities rely on a strict partitioning of cache resources. Thus, in contrast to related work, which rely on \CHANGED{probabilistic defenses (e.g., randomized cache line mappings~\cite{Trilla18,Qureshi18,ceaser-s,scattercache,phantomcache}), \sys provides certainty that the attacker cannot infer the cache accesses of a victim, if the partitioning is correctly implemented.} 
The main security goals of \sys are to prevent an adversary from accessing (read/write) data in the exclusive cache chunk of an \sd and to prevent eviction interference between the adversary and victim domain. In the following, we show how \sys achieves these goals with strict cache partitioning and we discuss why \sys's security guarantees even hold in the event of a strong adversary that compromised the operating system kernel.
Besides these security considerations, we verified the correctness of our implemented \sys prototype by explicitly issuing memory requests which try to read, write and evict cached data of \sd{}s. 

\paragraph{Strict Partitioning of \sd Cache Chunks} As described in~\Cref{sec:design}, the trusted software component communicates the number of chunk sets which should be assigned to an \sd to the \sys cache controller which configures the \etable and verifies that each cache chunk set is only assigned to a single \sd. At every cache memory access, the cache controller uses the domain ID to index the \etable and to retrieve the list of assigned sets (\textsc{SID-VEC}). Since the assignment of domain IDs and configuration of the \etable can only be performed by the trusted software component, the indexing logic of the cache controller will never return a cache set which does not belong to the issuer of the memory request. Thus, an adversary is never able to read an \sd{}'s exclusive sets (\dcache), write to them or evict them. As a result, \sys protects from access-based attacks, which require the adversary to flush memory out of the victim's sets, and conflict-based attacks, which require to fill the victim's sets and thus, evicting its cache lines. Moreover, \sys{}'s strict cache resource separation prevents an adversary from observing evictions of its own sets caused by the victim, which protects from occupancy-based attacks, and also strictly prevents the sharing of replacement policy metadata, which has been shown exploitable~\cite{Kiriansky17}. In general, the adversary can only infer how many cache sets are assigned to an \sd but cannot infer which sets (and therefore which memory addresses) are accessed at which point in time. 
\CHANGED{As described in~\Cref{subsec:attacker_model}, collision-based attacks are not considered. Defending against them architecturally requires locking the victim cache lines. \sys could be extended to integrate this, though mitigating an attack which is very specific to particular software implementations and is not widely shown does not justify the resulting large performance overhead.}

\sys allows for a dynamic assignment of cache sets to \sd{}s. Whenever the \dcache capacity of an \sd is modified, all assigned chunk sets are invalidated. This prevents leakage of sensitive \sd data when chunk sets are reassigned to another execution domain, and prevents an adversary from injecting malicious data into a set, when additional sets are assigned to an \sd. \CHANGED{The invalidation is however only required for the \sd whose \dcache is resized; all other \sds do not need to be modified and thus, their cache lines do not need to be flushed. The same applies when the \dcache for an \sd is completely de-allocated.}
An adversary could also try to trick an \sd into storing sensitive data in a mainstream cache line that is accessible for the adversary (\shared flag bit set). \sys prevents this by checking the metadata on every memory request of an \sd to verify that the memory region was indeed configured as \textit{shared}.

\paragraph{Protecting from Compromised \nsd} As described in~\Cref{subsec:attacker_model}, in the adversary model of TEE architectures, the OS (and therefore the \nsd) is not trusted, allowing an adversary to map physical memory pages of a victim \sd to its own memory space and to directly access it in the cache. If an \sd (represented by an enclave) demands side-channel protection (\exclusivemode), all data is cached in the exclusive \dcache and thus, not accessible for the adversary. However, if an \sd is not concerned about cache side channels (\mainstreammode), the data is cached in the shared mainstream sets and thus, must still be protected from malicious direct accesses. \sys prevents those attacks with the domain ID tag which is added to every cache line. On every cache write, the domain ID tag is set to the ID of the write request issuer. Subsequently, on every read request, the ID of the issuer is compared to the stored ID and the request only permitted if both IDs match. Evictions are permitted for every domain to achieve a perfect utilization of the shared cache sets. This is, however, not a security concern since an \sd{}'s data will only be cached in the shared sets if the \sd is in \mainstreammode or if the data is explicitly shared with the \nsd.

%% file: eval.tex
\section{Implementation \& Evaluation}
\label{sec:eval}
To evaluate \sys with respect to its hardware footprint, power consumption overheads, and performance impact, we implemented our design in hardware and on an architectural cycle-accurate simulator.

\paragraph{Methodology}
We implemented a hardware RTL model of \sys to extend an open-source RISC-V processor and synthesized it to evaluate the storage and logic overhead incurred. We use our hardware implementation to extract the additional cycle latencies incurred by \sys due to individual cache management and access operations. Then, to evaluate the performance impact of \sys on large mixed workloads, we extend an architectural cycle-accurate simulator, the gem5 simulator, with \sys and configure it to model a multi-core architecture with a 3-level cache hierarchy which matches our system assumptions (\Cref{subsec:system_model}). We incorporate the cycle latencies derived from our hardware implementation into our gem5 setup and use it to collect performance measurements on the standard SPEC CPU2017~\cite{spec_2017} benchmarks suite (aligned with related work~~\cite{Qureshi18,ceaser-s,scattercache,phantomcache}) to evaluate the overall performance impact of \sys. Complementary to the compute-intensive SPEC benchmarks, we also evaluate \sys on the I/O-intensive webserver \texttt{nginx}.
In order to achieve the most realistic results, we conduct our experiments in the full-system simulation mode of gem5 which simulates the user- and kernel-space software and also I/O devices.

We describe next our hardware implementation (\Cref{subsec:hw-impl}), performance evaluation (\Cref{subsec:perf}), and our hardware an power overhead evaluation (\Cref{subsec:hw-eval}).

\subsection{Hardware Implementation}\label{subsec:hw-impl}
In our hardware model, we extended the cache tag store with a 4-bit \idid and a 1-bit \shared bit to tag the owner domain of each cache line and whether it is shared with the \nsd (OS), respectively. We also extended the cache controller with the table structures shown in~\Cref{fig:tables}. To track the status of the 16,384 sets of a 16~MB LLC with 16-ways, the \cst is implemented as a 16,384-bit register that is indexed by the set ID to read out the corresponding 1-bit \textsc{ALLOC} flag. 
To support set allocation for 16 domains in parallel, the \etable is implemented as a 16-row \idid-indexed vector structure.
We decided for 16 parallel domains in our hardware implementation since this is also the maximum number of enclaves supported by multiple TEE architectures in parallel~\cite{keystone,bahmani2020cure}.
We define for our implementation that the maximum number of sets that can be allocated to any domain is 8,192 sets. Thus, we reserve 4 bits to represent the set \textsc{INDEX} bits number (to index into one of 8,192 sets), 114,688 bits (8,192 sets $\times$ 14 bits to represent each set's global ID) for the \textsc{SID-VEC}, and 1 bit \textsc{ALLOC} flag per domain. We discuss the storage overheads incurred by the tables in~\Cref{subsec:hw-eval}.

We implement the control finite-state-machines (FSMs) that receive cache allocation and de-allocation requests and perform the necessary management. For allocation, the FSM controls cycling through the sets sequentially to allocate free ones to the requesting \sd, updating their status in the \cst and updating the corresponding domain status in the \etable. 
For de-allocation, another FSM controls that the \textsc{SID-VEC} of the pertinent \sd is read from the \etable, its \textsc{ALLOC} flag reset, and then, all sets of that \sd de-allocated (by sequentially indexing through the \cst with the respective set IDs from the \textsc{SID-VEC}). Both allocation and de-allocation occur in powers-of-2 set numbers in our prototype. This is only an implementation decision in our prototype to minimize the logic complexity and overhead.

The cache access mechanisms are extended to include the \etable lookup required for \sys to identify which global set IDs belong to the issuing domain and to map the access to the correct set prior to tag lookup. Additionally, for \nsd accesses, after mapping to the correct set ID, concurrent sets
are computed and looked up in the \cst in parallel to identify which ones are unallocated.

\subsection{Performance Evaluation} \label{subsec:perf}
In this section, we first describe the latencies from our RTL model which we incorporate into our gem5 implementation. Next, we provide an evaluation of \sys's performance impact using the gem5 implementation.

\paragraph{Cycle Latencies}
\label{subsubsec:eval_cycles}
As described in~\Cref{sec:design}, \sys introduces a new indexing policy. For \sd memory requests in \exclusivemode, a lookup in the \etable is required. For requests in \mainstreammode and all \nsd (OS) memory requests, the mainstream sets must be looked up. The comparison of the stored \idid with the requester \idid is done in parallel with the address tag comparison and thus, does not introduce additional latency. For \sd requests in \exclusivemode, we measure an additional latency of 1 cycle and for \nsd requests and \sd requests in \mainstreammode of an additional 2 cycles.
For the access latencies of modern LLCs on multi-core systems, we estimate a baseline of 80 cycles in line with vendor multi-core processors~\cite{latencies_l3}. 

Whenever an \sd gets sets allocated, unallocated sets are looked up and the \etable updated. At de-allocation, sets of the \sd must be invalidated (and possibly flushed) 
and the \cst and \etable updated. 
For allocation, the overall latency incurred is variable and is a function of: 1.)~how many sets \textsc{CH-NUM} are requested for allocation, and 2.)~how many sets have to be looked up in the \cst. At the worst case, 
this incurs a latency of 16,384 cycles and at the best case, \textsc{CH-NUM} cycles. An additional 1 cycle is incurred to update the \etable subsequently. The \textsc{INDEX} is computed and communicated already by the trusted component in the allocation request, thus it does not contribute additional latency. 

For de-allocation, we measure an overall latency 
of \textsc{CH-NUM} + 2 cycles, where 1 cycle is required to look up the \etable, and another cycle to update it, followed by \textsc{CH-NUM} cycles to de-allocate each set in the \cst. At worst case, a latency of 8,194 cycles is incurred (assuming a maximum of 8,192 sets per domain). However, de-allocating the sets in \cst is done in parallel to invalidating (and possibly flushing if dirty) the respective cache lines.

We emphasize that allocating new sets to any \sd does not require \CHANGED{invalidating or flushing any other sets of the \nsd{} or other \sds which would require re-caching them. This is one key design goal of \sys since it eliminates this performance overhead on other domains, particularly the \nsd{}.} 
The allocation of sets either happens only once during the \sd setup or occasionally when the number of assigned sets is modified at runtime which requires a context switch out of the \sd. 
The \sys allocation/de-allocation overheads induced remain negligible when compared with the general overheads of TEE architectures~\cite{costan2016sanctum,sanctuary,keystone,bahmani2020cure}. Therefore, we do not invest in increased logic complexity to optimize the cycle overheads incurred for allocation and de-allocation, since they are not in the critical path, i.e., LLC accesses.

\begin{table}[t]
	\footnotesize
	\begin{center}
		\begin{tabularx}{\columnwidth}{c|c|c|c|c}
			\multirow{2}{*}{\textbf{Parameters}} & \textbf{L1} & \multirow{2}{*}{\textbf{L2}} & \textbf{L3} & \textbf{L3} \\
			& \textbf{(I\&D)} & & \textbf{(gem5)} & \textbf{(\sys)} \\
			\hline
			\hline
			\multirow{2}{*}{size} & 64~KB &  \multirow{2}{*}{512~KB} & \multirow{2}{*}{16~MB} & \multirow{2}{*}{16~MB} \\
			& \& 32~KB & & & \\
			\hline
			\# of sets & 128 \& 64 & 512 & 16,384 & 16,384 \\
			\hline
			associativity &  8-way & 16-way & 16-way & 16-way \\
			\hline
			access latency &  \multirow{2}{*}{4} & \multirow{2}{*}{14} & \multirow{2}{*}{80} & \multirow{2}{*}{81 / 82}\\
			(in cycles) & & & & \\
			\hline
		\end{tabularx}
		\caption{\small Cache configuration on our gem5 evaluation setup with an inclusive 3-level cache hierarchy.}
		\vspace{-2em}
		\label{tab:cache_config}
	\end{center}
\end{table}

\paragraph{Mixed-Workload Cycle-Accurate Evaluation}
\label{subsubsec:eval_gem5}
We implement \sys on the cycle-accurate gem5 simulator and construct a multi-core system which resembles a modern computing system with an inclusive 3-level cache hierarchy. Each core has access to a core-exclusive L1 and L2 cache, and an L3 LLC shared among all cores. For the L1 and L2, we use the unmodified cache implementation provided by gem5, whereas we use our \sys implementation for the L3 cache. The configuration parameters of each cache level are shown in~\Cref{tab:cache_config}. We derive realistic values for the cache sizes, number of cache sets, associativity and access latency in line with modern caches. For the \sys L3 cache, we add our induced latencies collected from our hardware implementation. Constructing a gem5-based multi-core system with 3-level cache hierarchy in full-system simulation mode to collect representative cycle-accurate traces for large workloads involved significant engineering challenges, as also evident by recent works that rely on trace-based simulators for their evaluation with SPEC workloads~\cite{Qureshi18,ceaser-s,scattercache,phantomcache}. 

\begin{figure}[h]
	\centering
	\includegraphics[width=1\columnwidth]{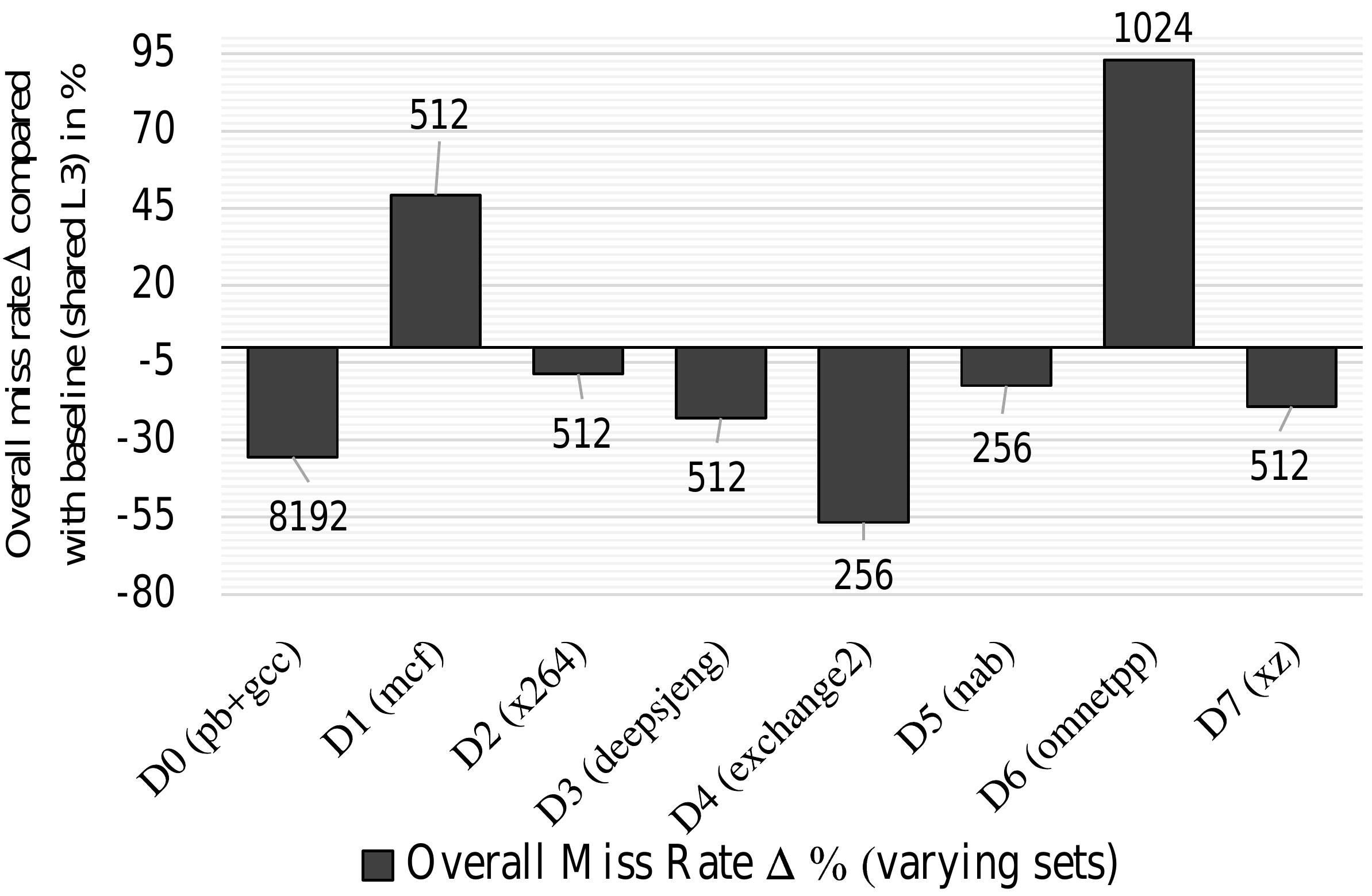}
	\caption{\small Cache miss rate impact of \sys for SPEC benchmarks on a 8-domain setup; compared to a shared L3 cache.}
	\label{fig:8-core}
\end{figure}

We measure the performance impact of \sys on real-world workloads by using the standard SPEC CPU2017 benchmarks with both the SPECspeed 2017 Integer and SPECspeed 2017 Floating Point suites which represent a wide range of compute-intensive applications such as compilers, video compression, machine learning or modeling tasks. \CHANGED{Since running all of the benchmarks on our full-system cycle-accurate gem5-based simulation setup would be very costly in terms of memory and time, we selected benchmarks from the different application domains and different working set sizes, guided by this memory-centric characterization of the SPEC CPU2017 benchmarks~\cite{spec-study}.} Moreover, to also cover I/O-intensive workloads, we evaluate the impact of \sys on the widely used webserver \texttt{nginx}.
We run our experiments for 1 trillion instructions before we start to collect measurements, in order to boot the system, start the benchmarks and collect more representative metrics. We run all our experiments for a total of 1 billion instructions in the full-system mode of gem5 and collect statistics to compute the Cycles Per Instruction (CPI) metric, in order to capture the additional latency effect, and the L3 cache miss rates for the reduced cache capacity effects. If not stated otherwise for single experiments, the miss rates are calculated as the geometric mean over the instruction and data miss rates of the page table walker and core. We compare \sys to 1.) a baseline system with an unmodified insecure L3 cache and to 2.) an L3 cache which implements a way-based partitioning scheme in which cache ways are assigned to \sd{}s as provided, e.g., by CATalyst~\cite{Liu16} which uses Intel CAT~\cite{intel_cat}, SecDCP~\cite{Wang16}, DAWG~\cite{Kiriansky17}, Keystone~\cite{keystone} or CURE~\cite{bahmani2020cure}.    
We evaluate \sys with a set of experiments which investigate different computing scenarios. First, we show how \sys's partitioning influences the performance of mixed workloads when encapsulated in \sds (in \exclusivemode). Then, we evaluate \sys's impact on the \nsd{} (OS-domain) and compare against way-based partitioned cache schemes. We conclude our evaluation with a set of experiments which show the scalability of \sys.
\CHANGED{In general, when comparing to the baseline (unpartitioned L3 cache shared by the same workload), our experiments show a negative effect of \sys on the performance of a benchmark when only a small \dcache size is assigned to it. However, when increasing the \dcache size this effect vanishes. At some point, depending on the specific characteristics of a benchmark, the exclusive \dcache assigned by \sys leads to a positive effect on its performance as we show in the following experiments. This gives the developer some degree of freedom to calibrate the performance of the workload by distributing the cache resources accordingly, e.g., to optimize the performance of a particular benchmark if desired given that the cache resources are available/affordable.} 
All experiments were conducted on an x86 platform equipped with an Intel Xeon Silver 4215 CPU (2.50 GHz) and 186 GB RAM.

\begin{figure}[t!]
	\centering
	\includegraphics[width=1\columnwidth]{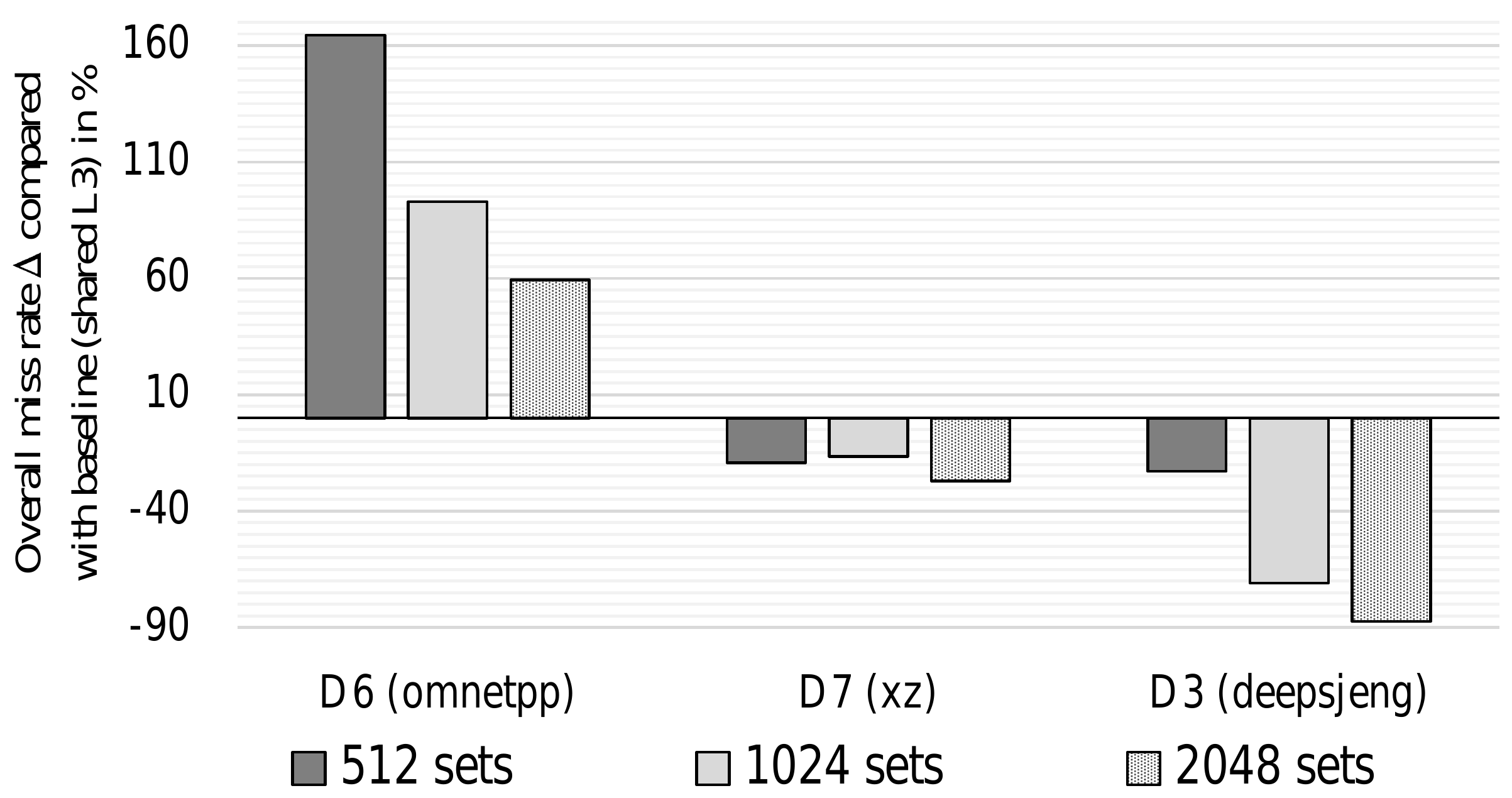}
	\caption{\small Cache miss rate impact of \sys for SPEC CPU2017 benchmarks (varying sets); compared to a shared L3 cache.}
	\label{fig:8-core-missrate}
\end{figure}

\begin{figure}[ht]
	\centering
	\includegraphics[width=1\columnwidth]{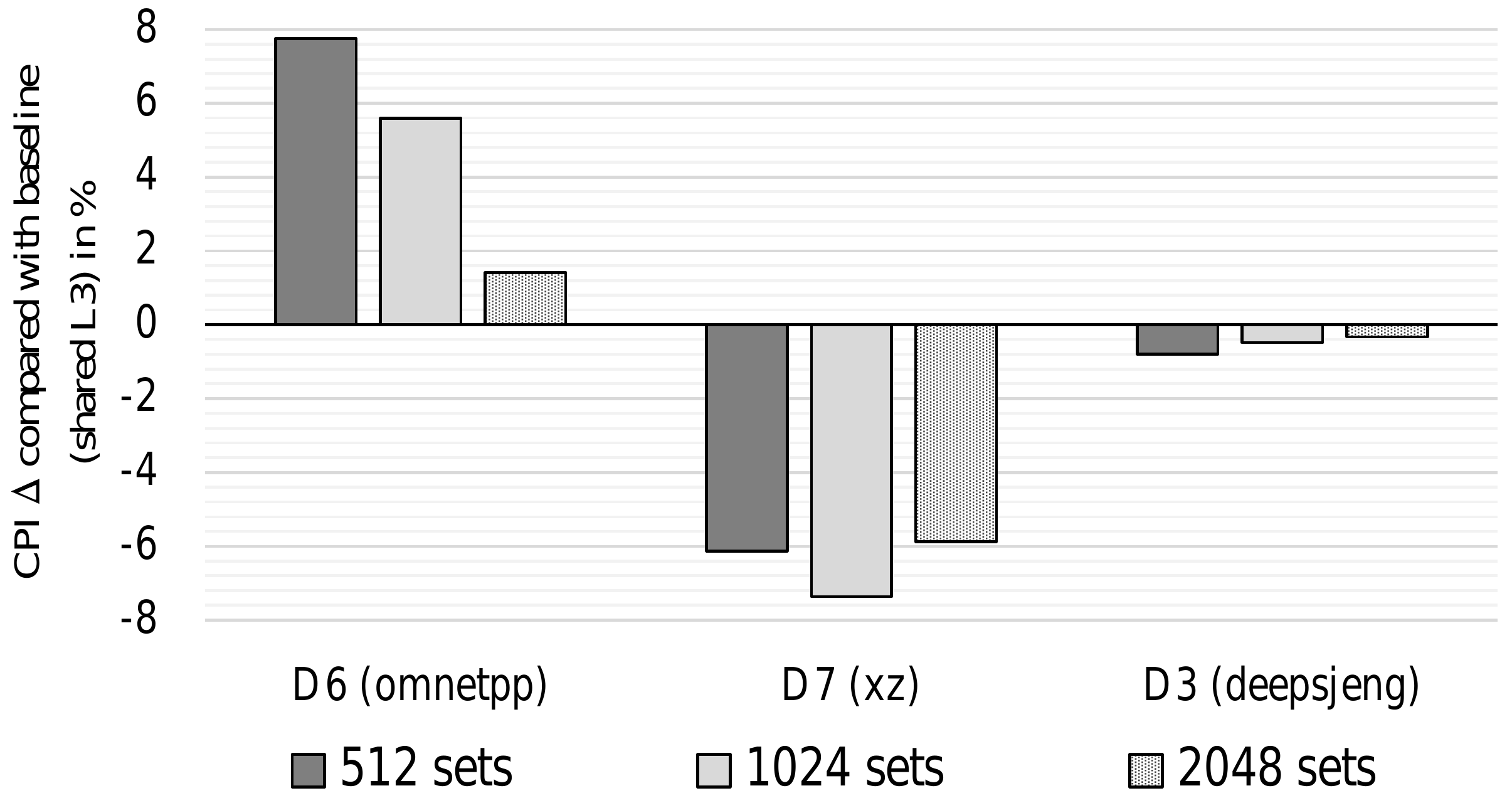}
	\caption{\small CPI impact of \sys for SPEC CPU2017 benchmarks (increasing sets); compared to a shared L3 cache.}
	\label{fig:8-core-cpi}
\end{figure}

\paragraph{\sd Performance Impact}
In the first set of experiments, we evaluate the performance impact \sys has on mixed workloads when protected in \sds in \exclusivemode. We run 7 randomly selected SPEC benchmarks in \sds and show our results in~\Cref{fig:8-core}. %
The \nsd (D0) runs Linux (kernel version 4.19.83) and 2 benchmarks with large working sets (\texttt{600.perlbench\_s} and \texttt{602.gcc\_s}).
In this experiment, we assign 8,192 sets to the \nsd and a varying number of sets to each \sd as indicated in the plot. \CHANGED{We chose the number of sets by briefly analyzing the working set size of the benchmark running in each \sd, and assigning bigger working sets to more cache sets. This is only required when optimizing for performance, otherwise a default number of sets can be assigned to each benchmark.} We observe in the experiment that the overall miss rate significantly decreases for most benchmarks when compared to sharing the L3 cache. This shows that the assignment of a smaller but exclusive cache portion can even reduce the cache miss rates of a workload. Moreover, our results indicate that the number of cache sets required to reduce or completely avoid the impact of \sys heavily depends on the characteristics of the workload. In our experiment, the benchmarks \texttt{605.mcf\_s} and \texttt{620.omnetpp\_s} would require more cache sets than the assigned 512 and 1024 sets to avoid an impact on the cache miss rates. 
We investigate this in another experiment where we customize the number of sets allocated to an \sd for some of the benchmarks and show how the miss rate decreases significantly when increasing the chunk size (\Cref{fig:8-core-missrate}). In another experiment (\Cref{fig:8-core-cpi}), we show how the varying chunk sizes also influence the CPI values. As for the miss rates, the CPI decreases in general. We observe, however, some outliers with the CPI metrics collected, owing to the complexity of a full-system multi-core simulation on gem5 which also includes unpredictable kernel runtime behavior into the statistics.

\begin{figure}[ht]
	\centering
	\includegraphics[width=1\columnwidth]{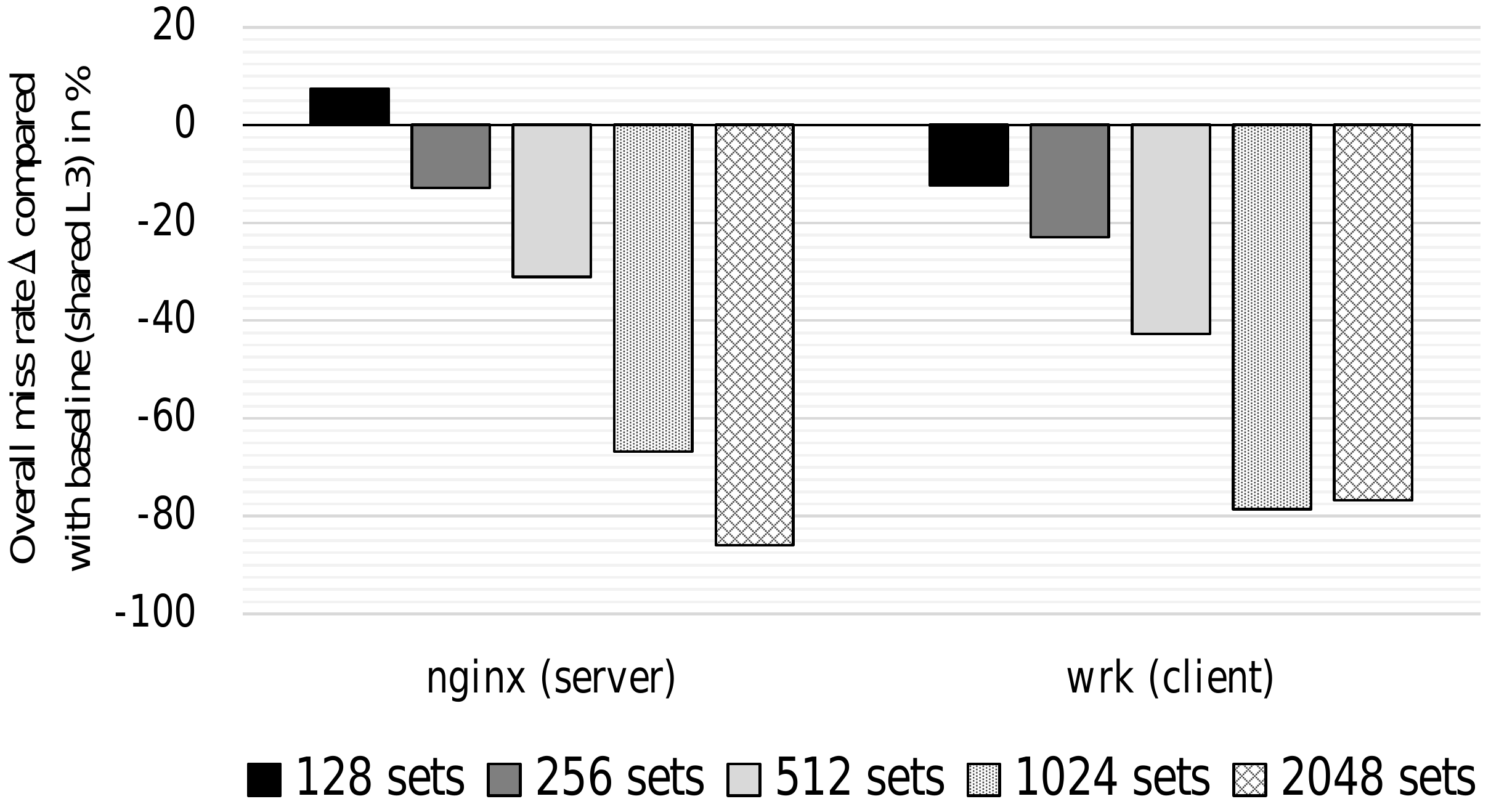}
	\caption{\small Cache miss rate impact of \sys for \texttt{nginx} and \texttt{wrk} (increasing sets); compared to a shared L3 cache.}
	\label{fig:nginx}
\end{figure}

Additionally, to  evaluate the impact of \sys on I/O-intensive workloads, we conduct experiments in which we run the \texttt{nginx} webserver in one \sd and the HTTP benchmarking tool \texttt{wrk} in another \sd, whereas we keep the \nsd unmodified. We then use \texttt{wrk} to send HTTP requests to the webserver using 12 threads and 400 open connections. In~\Cref{fig:nginx}, the miss rate impact of \sys on \texttt{nginx} and \texttt{wrk} is shown when increasing the number of sets from 128 to 2048. The results show, in line with our results on SPEC, how the increase of cache sets leads to a decrease in the overall miss rate. The decrease is already noticeable for a relatively small number of sets since the exclusive assignment of the cache sets prevents \texttt{nginx} and \texttt{wrk} from evicting the sets from one another.

\begin{figure}[ht]
	\centering
	\includegraphics[width=1\columnwidth]{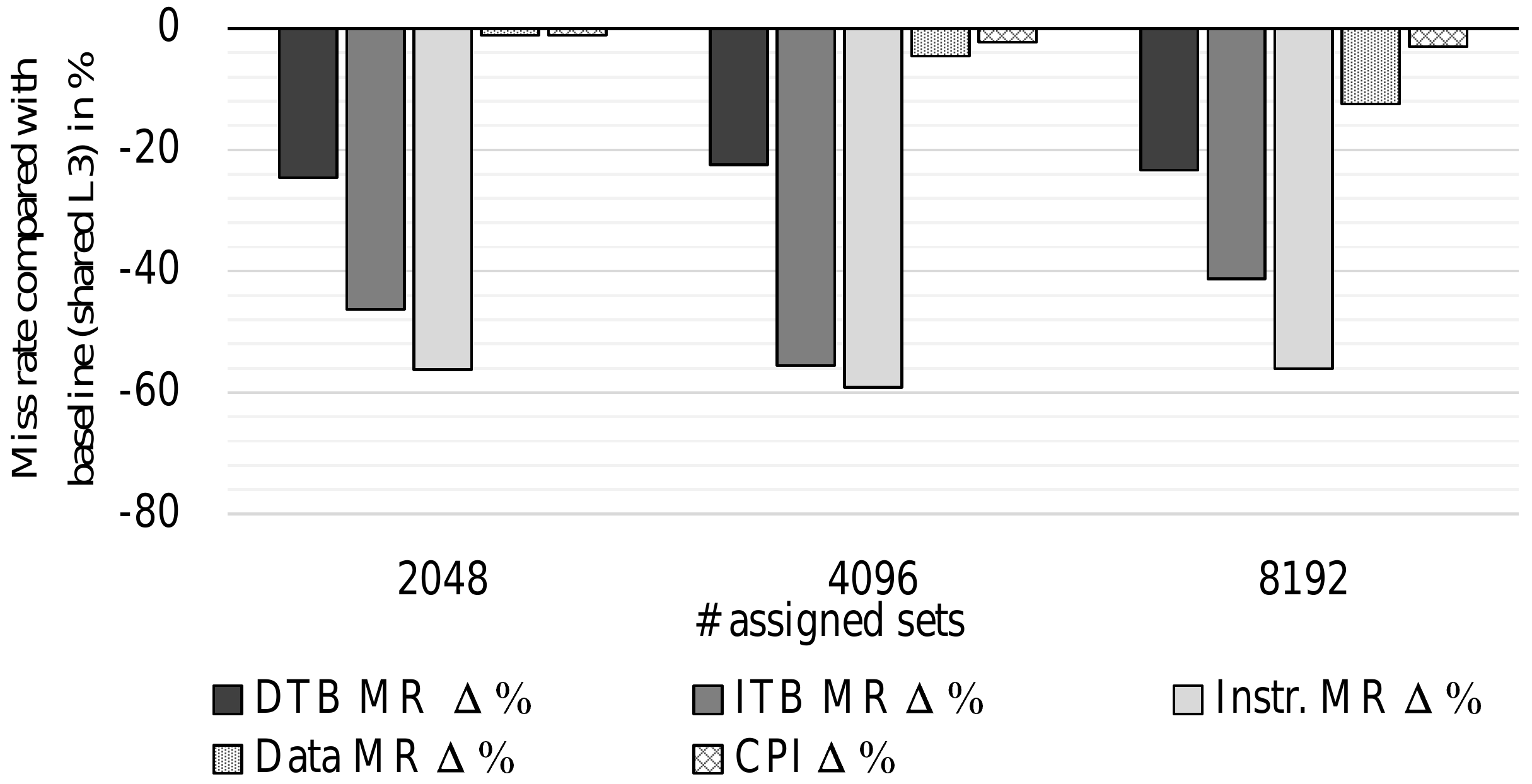}
	\caption{\small Miss rate \& CPI impact of \sys on the \nsd (increasing sets); compared to a shared L3 cache.}
	\label{fig:os-impact1}
\end{figure}

\paragraph{\nsd Performance Impact}
In the second set of experiments, we focus on the performance impact of \sys on workloads executing in the \nsd{}.
We again run mixed workloads from the SPEC benchmarks in \sds, while running Linux and the 2 memory-intensive benchmarks \texttt{600.perlbench\_s} and \texttt{602.gcc\_s} in the \nsd. 
In~\Cref{fig:os-impact1}, we vary the number of sets allocated to the \nsd from 2,084 to 8,192 while keeping the sets for the other domains unchanged. For these experiments, we show all 4 miss rate metrics over which we average in the other experiments, the data and instruction miss rates of the page table walker (DTB MR and ITB MR, respectively), and the data and instruction miss rates of the core (Data MR and Instr. MR, respectively).
While in general, all miss rates and CPI metrics decrease compared to the baseline, we only observe a slight improvement when increasing the chunk size from 2,084 to 4,096 and 8,192 sets.
This is because even when the number of statically allocated sets to the \nsd is rather small, the unallocated sets in the system (mainstream sets) remain available for the \nsd. Thus, performance is not significantly impacted for the \nsd and maximum utilization of the available resources \CHANGED{(as in an unmodified insecure cache architecture)} is preserved which was one of the key design goals of \sys. 

\begin{figure}[hb]
	\centering
	\includegraphics[width=1\columnwidth]{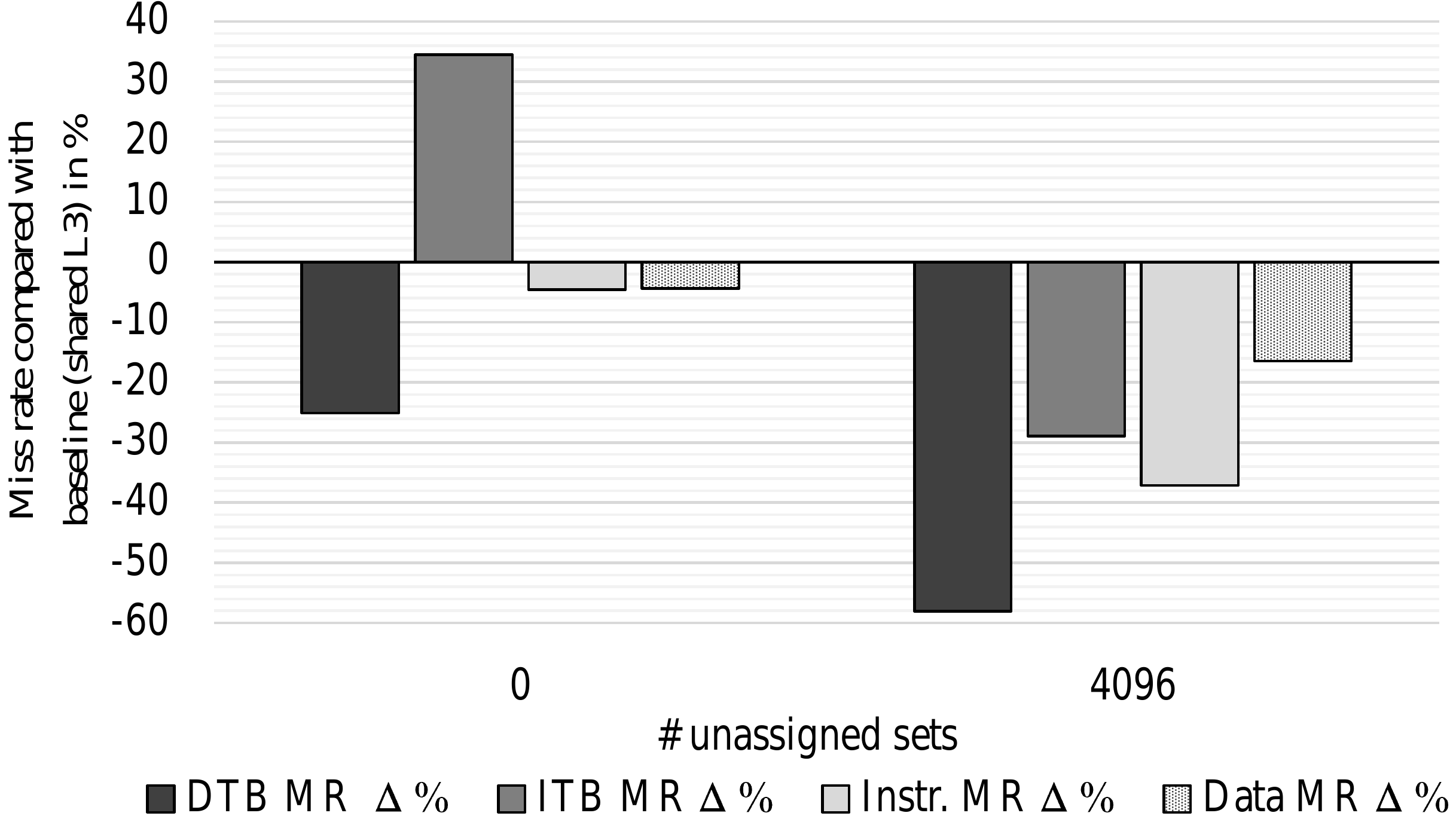}
	\caption{\small Miss rate of \sys on the \nsd with varying number of unassigned sets; compared to a shared L3 cache.}
	\label{fig:os-impact2}
\end{figure}

To investigate this, we run experiments (same setup) in which we assign 1,024 sets to the \nsd and vary the number of unassigned sets.
In the first run, all cache sets are allocated in our system, while in the second run, 4,096 sets remain unallocated and available for the \nsd. \Cref{fig:os-impact2} shows how the miss rates significantly decrease when 4,096 sets remain unallocated which demonstrates how \sys enables the \nsd to utilize unused cache sets.

\begin{figure}[ht]
    \centering
    \includegraphics[width=1\columnwidth]{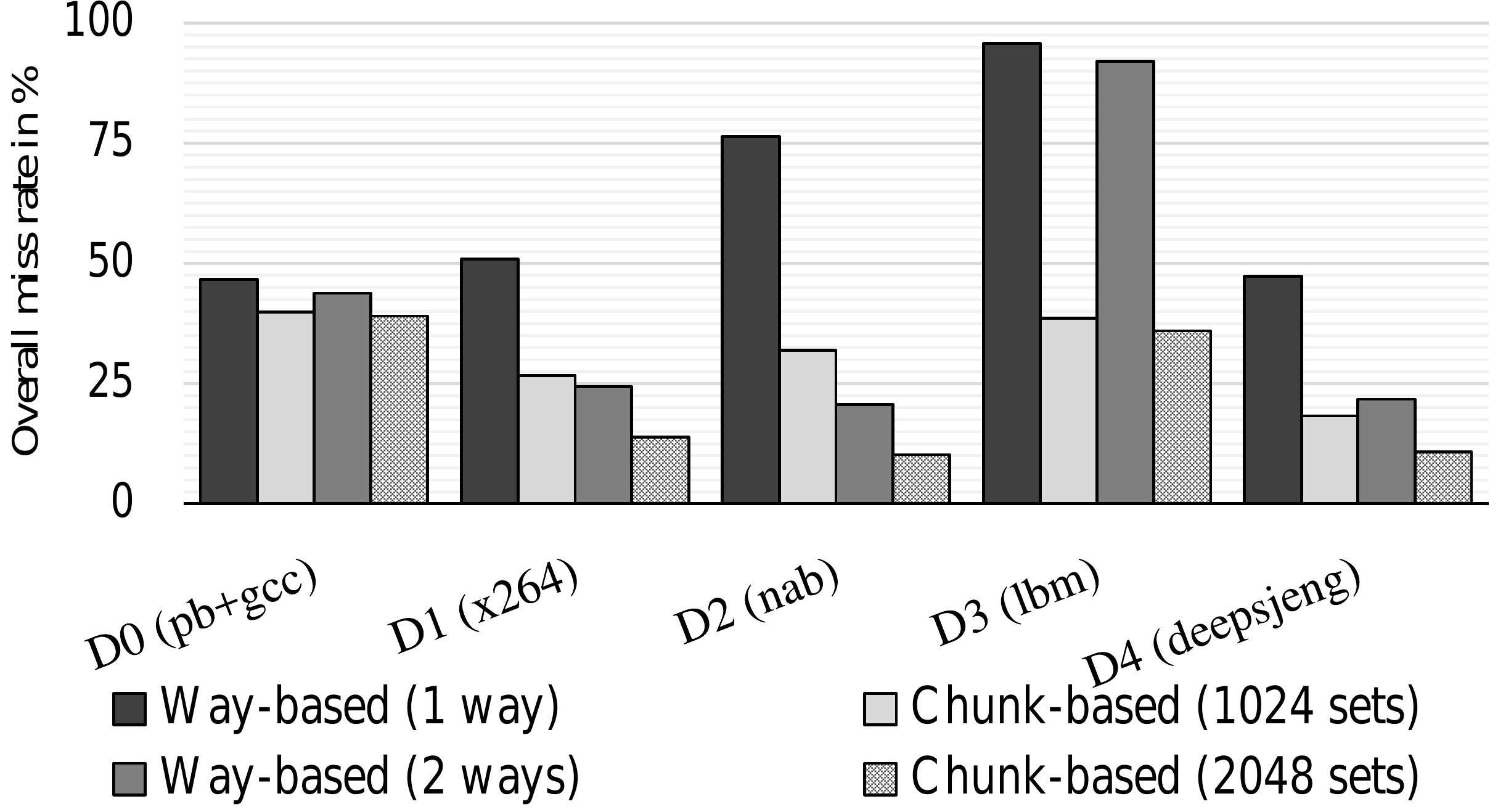}
    \caption{\small Overall miss rate for SPEC CPU2017 benchmarks with \sys; compared to a way-partitioned cache.}
    \label{fig:way-based}
\end{figure}

\CHANGED{\paragraph{Comparison with Partitioning-based Schemes}}
We compare \sys to a \CHANGED{cache partitioning scheme which we implement on gem5, specifically way-based partitioning,  being the only other strict cache partitioning approach.} We run a number of experiments with a 5-domain setup where we assign the same cache capacity to the same benchmark in both, the \sys and way-partitioned cache -- 1,024 or 2,048 sets in \sys and equivalently 1 way or 2 ways, respectively, in the way-partitioned setup. We show in~\Cref{fig:way-based} how for the same cache capacity, \sys outperforms way-based partitioning for randomly selected benchmarks. In fact, for some benchmarks such as \texttt{625.x264\_s} and \texttt{644.nab\_s}, allocating 1,024 sets even outperforms 2 ways (double the cache capacity) on a way-partitioned cache. We calculate an average decrease of 43\% in the miss rate for \sys vs. the way-partitioned cache for a 1~MB cache capacity (1024 sets) and a 39\% decrease for 2~MB (2048 sets).

\paragraph{Scalability and Dynamic Cache Allocation}
In~\Cref{app:eval}, we additionally evaluate \sys's ability to scale and support 32 \sds in parallel without degrading the performance of the \nsd (OS) and we also demonstrate how \sys supports the dynamic allocation of cache sets to an \sd during runtime.

\subsection{Hardware Footprint and Power Consumption Evaluation} \label{subsec:hw-eval}
To evaluate the storage and logic overhead incurred by \sys, we synthesize our implementation targeting a single-issue single-core RISC-V processor~\cite{roa} using Xilinx Vivado tools. While this processor does not provide an LLC, this is not necessary since we can still extend the existing simple cache controller to implement \sys, verify its functionality in cycle-accurate RTL-level simulations and evaluate its overheads.

\paragraph{Storage/Memory Overhead}
The main contribution to the hardware area overhead of \sys is the extra storage required, rather than the logic itself, since that requires the fabrication of memory which consumes more gates than hardware logic. The extra storage is needed for the additional tag bits required per cache line (4-bit \idid and 1-bit \shared flag), the \cst and \etable.
In our current prototype implementation targeting 16 domains, 16~MB LLC with 16-ways and 16,384 sets, and an allowed maximum of 8,192 sets per domain, the \cst consumes 2~KB, the \etable $\approx$ 224~KB, and the additional tag storage 160~KB, totaling 386~KB. This amounts to a negligible 2.3\% storage overhead relative to a 16~MB LLC which would consume approximately an additional 2.7\% area in fabrication.

\CHANGED{The capacity of these tables and the consequent storage area overheads are directly impacted by how the various design/implementation trade-offs involved are configured in different implementations of \sys, namely 1)~the number of active parallel domains supported (overhead increases only linearly) 2.)~the total L3 cache capacity and its number of sets, and 3.)~the maximum number of sets that can be allocated to a domain. For example, to support 32 domains, one more tag store bit is required costing an additional 0.25KB, relative to the overhead incurred for 16 domains as described above. The \cst capacity is unaffected, while the \etable capacity doubles to 448KB. The power consumption (evaluated below) would increase proportionally.}

\paragraph{Logic Overhead}
\sys requires extra hardware logic for the FSMs that handle the cache de-/allocation, and look up the tables prior to cache accesses (\Cref{subsec:hw-impl}). We synthesize our hardware implementation using Xilinx Vivado targeting a ZedBoard Zynq-7000 FPGA board, and estimate a logic overhead of $\approx$ 1.6\% relative to the single-core RISC-V processor that we extend. This would diminish relative to a significantly more complex multi-billion-transistor processor with a 3-level cache hierarchy which is the intended platform for \sys. \CHANGED{Furthermore, this overhead does not increase as the number of domains supported by \sys increases.}

\paragraph{Power Consumption Overhead}
\CHANGED{We focus here on the power consumption overheads incurred by the extended tag store and \cst and \etable tables, since the extra hardware added is dominated by them, and they contribute the most to the additional power consumption (static and dynamic) overheads. Besides, power consumption by cache memories is significantly more than logic, and is usually the largest contributor to the total power consumed by a chip.} We estimate the power consumption overheads of \sys in 22nm technology using the CACTI-6.0 tool~\cite{cacti}. For a 16-way 16~MB cache with 64~B cache line size, the total leakage power increases from 5056.57~mW (baseline) to 5313.83~mW. The \cst and \etable incur an additional 365~mW, amounting to a total of 12.3\% increase in the LLC power consumption. To support OS-specific chunk set indexing, the power consumption increases accordingly. If 2 sets are looked up in parallel (when 8,192 sets are allocated to the OS), the penalty on power consumption is negligibly minimal. When 4 or 8 sets are looked up in parallel, the power consumption overhead additionally increases by 5.5\% and 27.1\% relative to the baseline of 5056.57~mW, respectively. Relative to the overall chip power consumption of modern multi-core processors (90-150W), the LLC power consumption increases incurred by \sys remain reasonable.

%% file: related.tex
\section{Related Work}
\label{sec:related}
We categorize cache side-channel defenses which tackle the problem directly in the cache into two broad classes: partitioning-based and randomization-based. %
We focus in this section only on the most relevant works to \sys, 
which all propose hardware changes at the cache architecture.

\vspace{-.4em}
\subsection{Partitioning-based Microarchitectures} \label{subsec:partitioning-rel}
The partitioning-based defenses most related to \sys propose new cache architectures that assign cache resources (cache lines or ways) exclusively to protected domains. The TEE architectures Keystone~\cite{keystone} and CURE~\cite{bahmani2020cure} implement way-based partitioning to assign cache ways exclusively to enclaves. SecDCP~\cite{Wang16} forms security classes of applications with similar security requirements and assigns cache ways to them. DAWG~\cite{Kiriansky17} provides way-based cache partitioning in the context of speculative execution attacks. The main limitation of way-based partitioning is its inability to support a large number of protected domains in parallel since even large LLCs only comprise a small number of cache ways (up to 16). Moreover, these defenses lead to cache underutilization when assigned cache ways are not evenly utilized by a protected domain since the unused cache lines are blocked for all other domains on the system.

\sys, besides other approaches~\cite{Wang07,hybcache}, is more flexible since it partitions the cache on a cache-line basis. %
PLcache~\cite{Wang07} assigns cache lines exclusively to processes which allows for a strict and fine-grained partitioning of cache resources. However, PLcache's strict isolation does not allow for caching data shared between processes and strongly impacts the overall system performance and fairness of the cache utilization. %
Moreover, PLcache does not protect against occupancy-based attacks since the adversary can still infer the victim's memory accesses by observing that the victim is unable to access/evict cache lines.
 
HybCache~\cite{hybcache} assigns cache ways to protected domains (or enclaves) by providing a fully-associative mapping with random replacement for the ways to overcome the cache underutilization problem of way-based partitioning schemes. In contrast to PLcache, HybCache assigns only a subset of the cache resources to the protected domains which can be reclaimed by non-sensitive domains and thus, a fairer cache utilization is achieved which does not heavily degrade the overall system performance.
However, HybCache does not scale practically with large LLCs since it would incur high power consumption overheads. Moreover, HybCache does not provide strong security guarantees against occupancy-based attacks since it does not enforce a strict partitioning.

\CHANGED{In memory page-coloring schemes~\cite{Godfrey03,Kim12,Zhang16,Liu16,costan2016sanctum}, the mapping from physical memory addresses to cache lines is utilized to ensure that the cache lines used from sensitive applications do not overlap. One problem with page-coloring is its high impact on the software memory layout. It cannot fully support DMA and requires modifying the memory management (OS or hypervisor). Moreover, the assignment of cache lines is static, i.e., modifying the number of assigned cache lines during runtime would require to alter the physical memory layout of the software which is highly impractical.} 

\sys, however, provides flexible cache-line partitioning that can scale to support a larger number of protection domains than the number of cache ways. It additionally overcomes the limitations of other cache-line partitioning techniques by providing support for shared memory and by scaling to large LLCs while still providing strict isolation. \CHANGED{In contrast to page coloring schemes, \sys does not influence the memory layout, is compatible with commodity memory management software, and allows dynamic modification of the chunk sizes during runtime.}

\vspace{-.4em}
\subsection{Cryptographic Randomization Defenses} %
\label{subsec:randomization-rel}
These randomization techniques attempt to avoid the storage overhead of large randomized mapping tables that are deployed by earlier defenses~\cite{Wang07,Newcache16,Liu14} by relying on cryptographic primitives to reproducibly generate the randomized mapping.
Time-Secure Cache~\cite{Trilla18} uses a set-associative cache indexed with a keyed function using the cache line address and process ID as its input. However, a weak low-entropy indexing function is used, thus, frequent re-keying and cache flushing must be performed which increases complexity and performance impact.

CEASER~\cite{Qureshi18} also uses a keyed indexing function but without process ID. It also requires frequent re-keying of its index derivation function and re-mapping to limit the time interval available for an attacker to reconstruct the eviction set.
Under a minimal eviction set construction algorithm of $\mathcal{O}(\emph{$E^2$})$ complexity, CEASER has been shown able to withstand attacks with a re-keying rate of 1\%.
However, under eviction set construction techniques with $\mathcal{O}(\emph{$E$})$ complexity~\cite{ceaser-s}, the re-keying rate needs to increase to 35\%-100\%, which incurs 
prohibitively high performance overheads.
To resist these improved attacks, a skewed variant of CEASER, CEASER-S~\cite{ceaser-s} was proposed that divides the cache ways into multiple partitions (skews), with different encryption keys used for each partition. A cache line maps to a different set in each partition, where one of the partitions is chosen randomly for the line placement, making the minimal eviction set construction more difficult. 

ScatterCache~\cite{scattercache} also uses keyed cryptographic indexing 
where cache set indexing is different and pseudo-random for every protected domain but consistent for any given key. Thus, re-keying is still required at time intervals to hinder the profiling and minimal eviction set construction efforts.

Phantom-Cache~\cite{phantomcache} relies on a set of hardware-efficient hash function and XOR operations to map a cache line to 1 of 8 randomly chosen sets in the cache, each with 16 ways, thus, increasing the associativity to 128. This requires accessing 128 locations on each cache access to check if an address is cached, resulting in a high power overhead of 67\%. 

Defenses based on cryptographic primitives have multiple weaknesses: 1.)~These defenses remain only as secure as the best/fastest known attack strategy/minimal set eviction construction algorithm~\cite{casa-micro20,purnal2019advanced} with no solid future-proof security guarantees. \CHANGED{In fact, a recent work~\cite{song-randomized} has further shown that other attack techniques and workarounds can be used to exploit certain flaws in ScatterCache and CEASER-S to completely undermine them and their security guarantees.} 2.)~Their promised security guarantees often rely on the alleged, yet not thoroughly investigated unpredictability of low-latency cryptographic primitives. The primitives deployed by CEASER, CEASER-S and ScatterCache have been shown vulnerable to cryptanalysis which enables the construction of eviction sets without even accessing memory~\cite{systematic-gruss,brutus}. Deploying primitives that resist formal cryptanalysis is also not practical since it would incur increased latency, thus, further degrading performance in the cache's critical path. 3.)~If the re-keying rate is increased to mitigate novel attacks, the induced performance overhead renders these defenses impractical. 

Mirage, a concurrent work, attempts to overcome the vulnerability to newer faster eviction-set construction algorithms, by eliminating set-associative eviction altogether~\cite{mirage}. However, besides still being vulnerable to occupancy-based attacks, Mirage does not support selectively enabling side-channel resilience only for execution domains that require it, thus, incurring a performance slowdown on the entire workload.

\sys, in contrast, eliminates the described unreliability and inflexibility fundamentally by providing strict, yet perfectly configurable and selective, partitioning across the execution domains. This enables each domain to allocate the cache capacity it requires and thus, experience the performance that it has opted to tolerate accordingly. This different paradigm provides well-grounded security assurances that stand the test of advances in cache side-channel attacks and different attack methodologies and complexities, without sacrificing performance. Instead, it provides by-design the possibility
to tune the security-performance trade-off for each domain as desired, without overtaxing the OS either.

%% file: conclusion.tex
\section{Conclusion}
\label{sec:conc}
In this paper, we presented a novel side-channel-resilient cache microarchitecture, \sys, for TEE architectures, that enables each execution domain to flexibly and selectively configure its exclusive cache sets only when cache isolation and side-channel resilience is required. Unlike randomization-based cache microarchitectures recently proposed, \sys fundamentally mitigates side-channel attacks by enforcing strict cache partitioning, thus providing future-proof and solid security guarantees. It also outperforms way-based partitioning and scales to support a larger number of execution domains, without degrading the performance of the OS. In this work, we show how \sys incorporates this configurable performance-security trade-off by design in the cache microarchitecture to cater most optimally for TEE architectures. Through our security analysis and evaluation, we also show how on-demand sophisticated side-channel security, as well as performance, functionality and usability requirements are preserved in \sys, \CHANGED{with small} hardware and memory costs.

%% file: app-eval.tex
\section{Additional Experiments}
\label{app:eval}

\subsection{\sd Scalability}
We also demonstrate how \sys scales for a larger number of parallel domains. As described in~\Cref{sec:design}, the design of \sys allows to support more domains in parallel than the 16 domains we choose for our hardware implementation. Thus, we conduct scaling experiments where we run the \texttt{619.lbm\_s} benchmark on every \sd and we increase the number of \sds from 4 to 8, 16 and up to 32. Running more \sds in parallel is not possible on our evaluation platform since the gem5 full-system simulation with 32 \sds already consumes the complete 186 GB of available RAM which unavoidably imposes certain limitations on our experiments. Given these constraints, we selected \texttt{619.lbm\_s} as a benchmark because of its relatively small working set. Throughout these experiments, the \nsd (which runs the Linux kernel and one instance of \texttt{619.lbm\_s}) gets 8,192 sets assigned. The overall miss rates for the \nsd, when scaling from 4 to 32 \sds, are stable, reaching 71.45\%, 71.64\%, 72.06\% and 71.75\%, respectively. Thus, with \sys, also a high number of \sds can be supported without degrading the performance of the \nsd (OS). Running even more domains was only limited by the memory constraints of our evaluation platform.

\subsection{Dynamic Set Allocation}
In another experiment, we analyze how the dynamic set allocation capabilities of \sys impact the \nsd and \sds during runtime. For this, we select a SPEC benchmark (\texttt{631.deepsjeng\_s}) which achieves a relatively small average cache miss rate, when enough cache sets are available, in order to better demonstrate the behavior of the dynamic set allocation. We run the benchmark in 4 distinct \sds and as part of the \nsd. We simulate 24 billion cycles on our evaluation platform which corresponds to 12s worth of computing (given that we simulate processors with a clock frequency of 2 GHz). At the beginning of the experiment, the \nsd (D0) gets 8,192 sets assigned, the \sds D1-D3 512 sets each and the \sd D4 only 1 set. Then, during runtime, the size of D4's chunk is modified. After 3s, the chunk size is increased to 512 sets, after 6s to 2048 sets and after 9s decreased to 1 set. The chunk sizes of the domains D0, D1, D2 and D3 are kept constant throughout the duration of the experiment. We collect miss rate statistics for all domains every 75ms (150,000,000 cycles) and compute the arithmetic mean over the instruction and data miss rates of the page table walker and core.

\begin{figure}[ht]
	\centering
	\includegraphics[width=1\columnwidth]{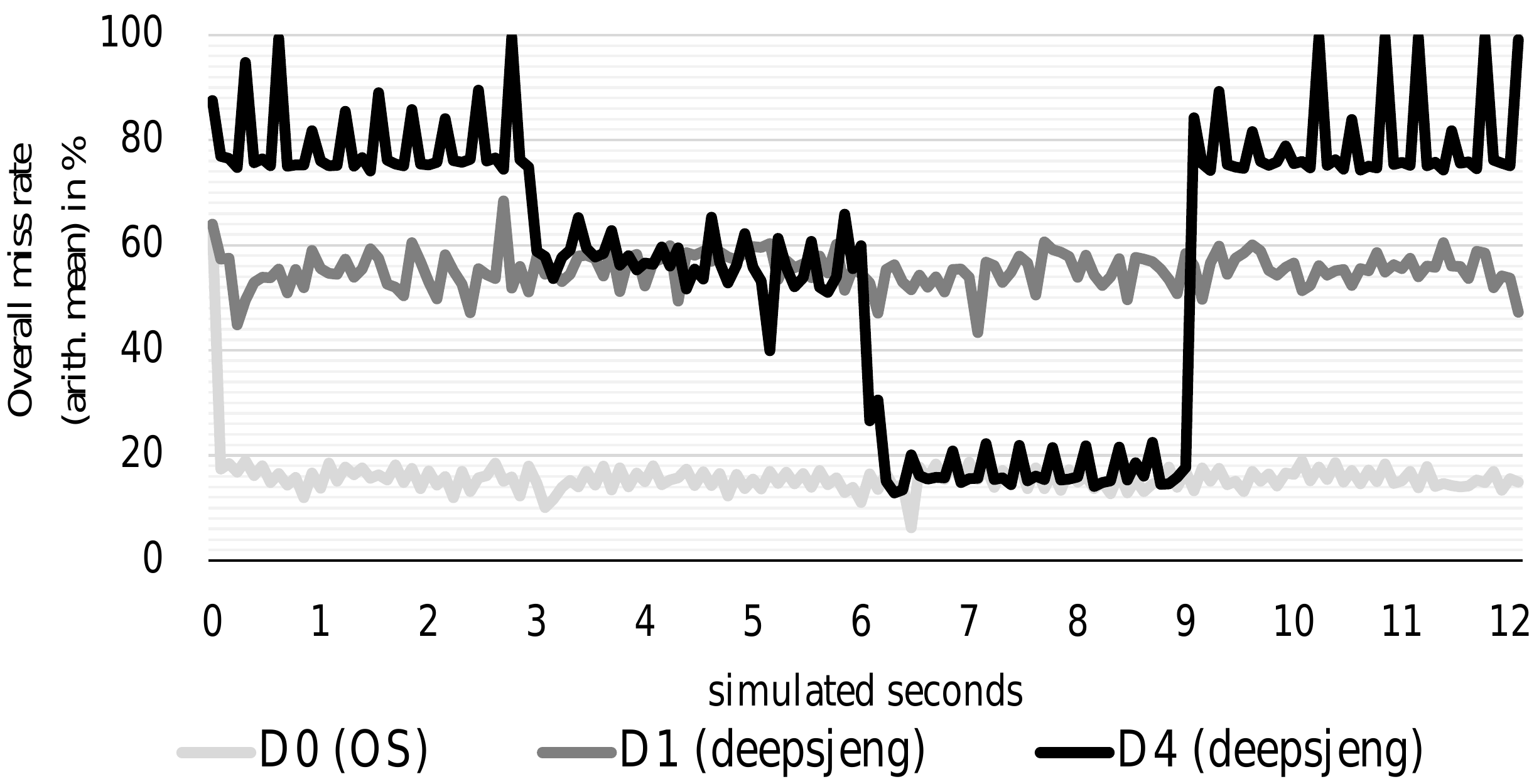}
	\caption{\small Cache miss rate impact of \sys on the \nsd (D0) and \sds (D1, D4) when dynamically modifying the size of the cache chunk assigned to D1.}
	\label{fig:dynamic_alloc}
\end{figure}

The results of the experiment are shown in~\Cref{fig:dynamic_alloc}, whereby we only show the miss rates for D0, D1 and D4 since the results of D2 and D3 are very similar to those of D1. The plot clearly shows how the increase and decrease of the chunks size affects the miss rate of D4. At the beginning, when only 1 set is assigned to D4, the miss rate fluctuates heavily around a value of 80\%. At the time point 3s, when 511 additional sets are assigned to D4, the miss rate almost immediately drops to around 60\%, thereby catching up with the miss rates achieved by D1. After another 3s, when D4's chunk size is increased to 2048, a low and stable miss rate of 20\% is achieved. The fact that D0 experiences the same miss rate with 8,192 sets shows that applications are not always benefiting from an increased chunk size and thus, available sets are better redistributed to other benefiting domains to improve the overall system performance. After 9s, the chunk size is decreased to 1 set which again leads to a heavily fluctuating miss rate of around 80\%.

Another interesting take-way from~\Cref{fig:dynamic_alloc} is that the flushing of all chunk sets, which happens after 6s, does not negatively influence the miss rate of D4, at least not when collecting the miss rate statistics at intervals of 75ms.

